\documentclass[12pt, draftclsnofoot, onecolumn]{IEEEtran}
\usepackage{geometry}
\ifCLASSINFOpdf

\else

\fi
\usepackage[cmex10]{amsmath}
\usepackage{amssymb}
\usepackage{cite}
\usepackage{graphicx}
\usepackage{array,color}
\usepackage{amsmath}
\usepackage{stfloats}
\usepackage{graphicx}
\usepackage{subfigure}
\usepackage{tabularx}
\usepackage{epsfig,epsf,color,balance,cite}
\usepackage{algorithmic}
\usepackage{algorithm}
\usepackage{bm}
\usepackage{textcomp}

\begin{document}

\title{Joint Pilot Allocation and Robust Transmission Design for Ultra-dense User-centric TDD C-RAN with Imperfect CSI}
\author{ Cunhua Pan, Hani Mehrpouyan, Yuanwei Liu, Maged Elkashlan, and Arumugam Nallanathan, \IEEEmembership{Fellow, IEEE}
%\thanks{This work was supported by ...}
\thanks{C. Pan and M. Elkashlan are with the Queen Mary University of London, London E1 4NS, U.K. (Email:\{c.pan, maged.elkashlan\}@qmul.ac.uk). H. Mehrpouyan is with the Department of Electrical and Computer Engineering, Boise State University, Boise, ID 83725 USA.  (e-mail: hani.mehr@ieee.org). Y. Liu and A. Nallanathan are with the Department of Informatics, King¡¯s College London, London WC2R 2LS, U.K. (e-mail:\{yuanwei.liu,arumugam.nallanathan\}@kcl.ac.uk). }
}

%\author{Cunhua Pan,  Huiling Zhu, Nathan J. Gomes and Jiangzhou Wang
%\thanks{This work is supported by  European Commission Horizon2020 project iCIRRUS under grant agreement No 644526. }
%\thanks{C. Pan, H. Zhu, N. Gomes and J. Wang are with the School of Engineering and Digital Arts, University of Kent, Canterbury, Kent, CT2 7NZ, U.K. (Email:\{C.Pan, H.Zhu, N.J.Gomes, J.Z.Wang\}@kent.ac.uk).}
%}

\maketitle
\vspace{-1.9cm}
\begin{abstract}
This paper considers the unavailability of complete channel state information (CSI) in ultra-dense  cloud radio access networks (C-RANs). The user-centric cluster is adopted to reduce the computational complexity, while the incomplete CSI is considered to reduce the heavy channel training overhead, where only large-scale inter-cluster CSI is available. Channel estimation for intra-cluster CSI is also considered, where we formulate a joint pilot allocation and  user equipment (UE) selection problem to maximize the number of admitted UEs with fixed number of pilots. A novel pilot allocation algorithm is proposed by considering the multi-UE pilot interference. Then, we consider robust beam-vector optimization problem subject to UEs' data rate requirements and fronthaul capacity constraints, where the channel estimation error and incomplete inter-cluster CSI are considered. The exact data rate is difficult to obtain in closed form, and instead we conservatively replace it with its lower-bound. The resulting problem is non-convex, combinatorial, and even infeasible. A practical algorithm, based on UE selection, successive convex approximation (SCA) and  semi-definite relaxation approach, is proposed to solve this problem with guaranteed convergence. We strictly prove that semidefinite relaxation is tight with probability 1. Finally, extensive simulation results are presented to show the fast convergence of our proposed algorithm and demonstrate its superiority over the existing algorithms.

\end{abstract}
%\begin{keywords}
%Ultra-dense networks (UDN), C-RAN, imperfect CSI, pilot reuse, virtual cell, fronthaul capacity constraints.
%\end{keywords}

\IEEEpeerreviewmaketitle
%\newpage
\section{Introduction}

The fifth-generation (5G) wireless system is excepted to offer a 1000x increase in capacity before 2020 to meet the ever-growing capacity demand in mobile wireless systems \cite{Andrews2014}. Cloud radio access network (C-RAN) is an emerging network architecture that shows significant promises in achieving this goal \cite{junwu2015}. A typical C-RAN has three main components: 1) Baseband unit (BBU) pool hosted in a cloud data center; 2) Radio remote heads (RRHs) geographically distributed over the coverage area; 3) High-bandwidth low-latency fronthaul links that connect the RRHs to the BBU pool.  The key feature of C-RAN is that the baseband signal processing in traditional base stations (BSs) are migrated to the BBU pool so that the conventional BSs can be replaced by low-functionality RRHs, which transmit/receive radio frequency signals. Due to their simple functionalities, RRHs can be densely deployed to provide ubiquitous service access for a large number of user equipments (UEs) in hot spots such as shopping malls, stadium, etc. Unlike the conventional ultra-dense small cells where cochannel interference (CCI) is a limiting factor \cite{shenchen2016}, under ultra-dense C-RAN architecture, centralized signal processing such as coordinated multi-point (CoMP) technique can be adopted to effectively mitigate the CCI thanks to the powerful computational capability at the BBU pool. Ultra-dense C-RANs are usually defined as those networks that there are more RRHs than the UEs \cite{daivd2015,Stefanatos2014,Kamel2016}, i.e., $\lambda_r>\lambda_u$, where $\lambda_r$ and $\lambda_u$ represent the density of RRHs and UEs, respectively. However, these works have not provided  quantitative values for those densities. In \cite{xiaohuge}, Ge \emph{et. al}  showed that the density of RRHs in 5G ultra-dense networks is expected to be up to 40-50 RRHs/${\rm{km}}^2$ to satisfy the seamless coverage requirements.

Recently, transmission design has been extensively studied to deal with various technical issues in conventional C-RAN such as reducing network power consumption in \cite{Yuanming2014}, or/and tackling limited fronthaul capacity constraints in \cite{Binbin2014}, or/and reducing the computational complexity \cite{pan2017joint} where user-centric cluster method is adopted. However, the most troublesome challenge is that dense C-RAN requires large amount of CSI to facilitate centralized signal processing. The acquisition of these CSI requires large amount of training overhead that increases with the network size. Results in \cite{Caire2010} showed that the network performance may even decrease with increasing number of RRHs when taking into account the cost of acquiring CSI. One promising way to reduce the training overhead is to consider the incomplete CSI case, where each UE only measures its CSI from the RRHs in its serving cluster (named intra-cluster CSI) and only tracks the large-scale fading (path loss and shadowing) for the CSI outside its cluster (named inter-cluster CSI).  Recently, transmission design considering the incomplete CSI has attracted extensive research interests \cite{Shi2014ICC,pan2017jsac,Lakshmana2016}.

However, intra-cluster CSI was assumed to be perfectly known in \cite{Shi2014ICC,pan2017jsac,Lakshmana2016}, which is impractical for dense C-RAN. The reasons are given as follows. To estimate the intra-cluster CSI in time-division duplex (TDD) C-RAN, the uplink training pilot sequences sent from the UEs that share at least one serving RRH should be mutually orthogonal so that the BBU pool can differentiate the CSI of the shared RRH to the corresponding UEs. One naive method is to assign all the UEs with mutually orthogonal pilot sequences. However, the number of time slots required for training will increase linearly with the number of UEs, which is unaffordable for ultra-dense C-RAN with large number of UEs. Hence, to support more UEs, one should allow some UEs to reuse the same pilots.  The pilot reuse scheme will incur the pilot contamination issue, which results in sizeable channel estimation error. Hence, in this paper, besides the incomplete inter-cluster CSI,  we also consider channel estimation procedure of intra-cluster CSI for a user-centric ultra-dense TDD C-RAN by designing pilot reuse scheme, and then design beam-vectors based on imperfect intra-cluster CSI. To the best of our knowledge, this paper is the first attempt to unify the  channel estimation and transmit beam-vector design for ultra-dense C-RAN into one general framework.

Pilot reuse design has been extensively studied in massive MIMO \cite{liyou15,Hyin13,Hyin2016,xzhu15,Zhao16}, where the proposed schemes are mainly based on the idea of assigning the same pilot to the UEs with different angles  due to the special feature of massive MIMO. Due to the limited number of antennas at the RRHs, the pilot reuse schemes designed for massive MIMO cannot be extended to the ultra-dense C-RAN network. Recently, \cite{Chen2016tvt} proposed a novel pilot allocation scheme for user-centric C-RAN based on the graph coloring algorithm such as the Dsatur algorithm \cite{brelaz1979new}. The aim of this algorithm is to minimize the number of required pilots for a given set of UEs under some practical constraints. However, for the sake of low implementation cost, the current LTE standards \cite{Stands} suggests that the proportion of pilots allocated to the UEs for training is fixed during one coherence time, i.e.,  where the training overhead of  $1\%$ is considered in LTE specification with a 10 ms training period \cite{Stands}. Hence, we should consider the joint pilot allocation and UE selection method to maximize the number of admitted UEs under fixed number of available pilots. In addition, the transmission beam-vector design is not considered in \cite{Chen2016tvt}. In this paper, we also consider robust beam-vector design under the pilot reuse scheme by taking into account the joint effects of pilot contamination and incomplete inter-cluster CSI. Unfortunately, due to the imperfect intra-cluster CSI, the beam-vector design based on the well-known weighted minimum mean square error (WMMSE) method \cite{Qingjiang2011}, that has been widely applied in the C-RAN setup \cite{Mingyi2013,Binbin2014,wcliao2014,XHuang16,pan2017joint,pan2017jsac} with perfect intra-cluster CSI, cannot be extended to solve the robust beam-vector optimization problem considered in this paper. Specifically, the contributions of this paper are given as follows:
\begin{enumerate}
  \item We  consider a two-stage optimization problem for dense C-RAN, i.e., channel estimation for intra-cluster CSI in Stage I and robust transmit beam-vector design in Stage II. Due to the constraint that the UEs served by at least one common RRH should be allocated with different pilots, some UEs may not be supported for given small number of available pilots $\tau$, considering large number of UEs in ultra-dense C-RAN. Hence, in Stage I, we formulate a joint UE selection and pilot allocation optimization problem to maximize the number of admitted UEs under several practical constraints. To solve this problem, we first  apply the Dsatur algorithm to find the minimum number of required pilots $n^*$ for supporting all UEs in the network. If $n^*>\tau$, one novel UE removement method is proposed by considering the multi-UE pilot interference. If $n^*<\tau$, all UEs can be admitted and a novel algorithm is proposed to fully reallocate all  available pilots to further reduce pilot contamination.
  \item Based on the results from Stage I,  in Stage II  transmit beam-vectors are designed to minimize the transmit power based on the imperfect intra-cluster CSI  and incomplete inter-cluster CSI. Both UEs' data rate and fronthaul capacity constraints are considered. It is observed that the expectation with respect to the channel estimation error and small-scale inter-cluster CSI makes it difficult to derive closed-form data rate expressions. To deal with this difficulty, we conservatively replace the data rate of each UE with its lower bound obtained by the Jensen's inequality. Furthermore, considering that the problem may be infeasible, we construct an alternative optimization problem that simultaneously maximizes the number of admitted UEs and minimizes the transmit power. Then, one iterative algorithm based on successive convex approximation (SCA) technique \cite{dinh2010local} and semi-definite relaxation approach \cite{Gershman2010} is proposed to solve this optimization problem with guaranteed convergence. We  prove that semidefinite relaxation is tight with probability 1.

\end{enumerate}

This paper is organized as follows. In Section \ref{system}, we introduce the signal transmission model along with the channel estimation for intra-cluster CSI. In Section \ref{stage1}, we propose the joint pilot allocation and UE selection algorithm. In Section \ref{userselectionalg}, we develop a robust transmit beam-vector optimization algorithm while taking the channel estimation error into account. Extensive simulation results are provided in Section \ref{simlresult}. Finally, this paper is concluded in Section \ref{conclu}.

\emph{Notations}: For a set ${\cal A}$, $\left| {\cal A} \right|$ is the cardinality of ${\cal A}$, while for a complex number $x$, $\left| x \right|$ represents the magnitude of $x$.  `s.t.' denotes `subject to'. ${{\mathbb{E}}_{\{ x\} }}\{ y\} $ denotes the expectation of $y$ over $x$. The complex Gaussian distribution is denoted as ${\cal C}{\cal N}(\cdot, \cdot)$ and  ${\mathbb{ C}}$ is used to represent the complex set. The lower-case bold letters means vectors and upper-case bold letters denote matrices.

\section{System Model}\label{system}
\vspace{-0.5cm}\subsection{Signal Transmission Model}

\begin{figure}
\centering
\includegraphics[width=3.2in]{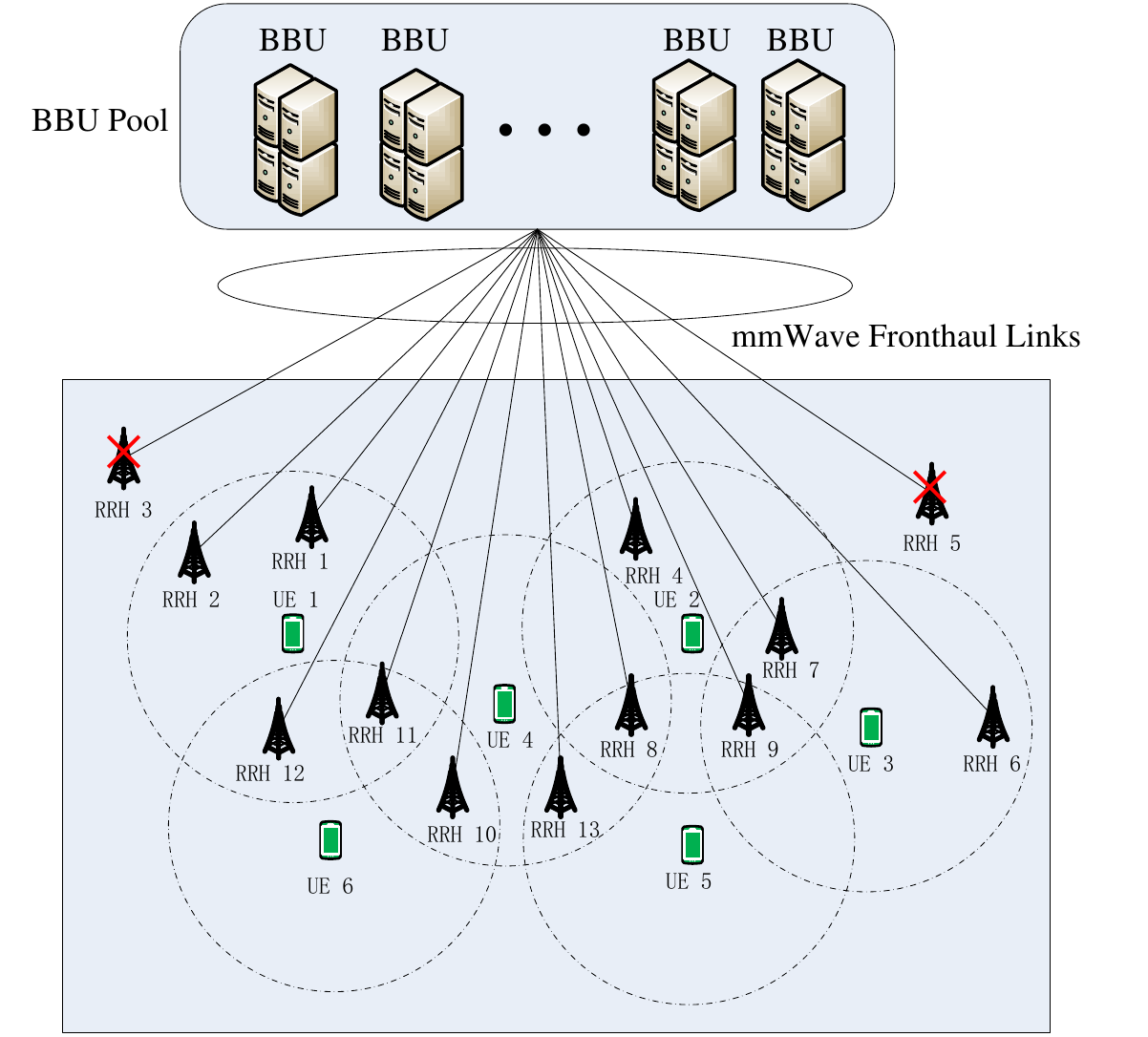}\vspace{-0.5cm}
\caption{Illustration of a C-RAN with thirteen RRHs and six UEs, i.e., $I=13$, $K=6$. To reduce the operation cost,  mmWave communication is used to establish the fronthaul link that connects each RRH to the BBU pool. To reduce the complexity, each UE is served by the RRHs within the dashed circle centered around the UE. }
\label{fig1}\vspace{-0.5cm}
\end{figure}

Consider a downlink dense TDD C-RAN with $I$ RRHs and $K$ UEs as shown in Fig.~\ref{fig1}.  Each RRH and each UE are equipped with $M$ transmit antennas and a single receive antenna, respectively. Denote the set of RRHs and UEs as ${\cal I} = \left\{ {1, \cdots ,I} \right\}$ and $\bar {\cal U} = \left\{ {1, \cdots ,K} \right\}$, respectively. Each RRH is connected to the BBU pool through the wireless fronthaul links employing mmWave communication technologies, which are represented by dark lines in Fig.~\ref{fig1}. The BBU pool is assumed to have all UEs' data and distribute each UE's data to a carefully selected set of RRHs through the wireless fronthaul links. We further assume that all the RRHs  transmit to the UEs with the carrier frequency below 6 GHz. As a result, the transmission from the BBU pool to the RRHs and the transmission from RRHs to the UEs can take place at the same time without interfering with each other.
% This paper focuses on the transmission from RRHs to UEs.
%The wireless fronthaul links can significantly reduce the operation cost and is much more flexible than the wired fronthaul links compared with optical fiber links \cite{Stephen2017}.

The set of UEs that are admitted in this network is denoted by ${{ {\cal U}}}\subseteq {\bar {\cal U}} $. To reduce the computational cost for dense C-RAN, the user-centric cluster method is adopted, e.g., each UE is only served by its nearby RRHs since distant RRHs contribute little to the UE's signal quality due to the severe path loss. Denote ${{\cal I}_k} \subseteq {\cal I}$ and ${{ {\cal U}}_i}\subseteq { {\cal U}} $ as the set of RRHs that potentially  serve UE $k$ and the set of UEs that are potentially served by RRH $i$, respectively.  Note that for ultra-dense C-RAN with many UEs and RRHs,  the clusters for the UEs may overlap with each other, i.e., there may exist two distinct UEs $k$ and $k'$ that ${{\cal I}_k}\cap {{\cal I}_{k'}}\ne \emptyset$, for  $\forall k,k'\in { {\cal U}}$. In practice, the clusters for each UE are formed based on the long term channel state information (CSI) such as large-scale fading \cite{LiuTVT2016}. Specifically, each UE measures  its average channel gains to the RRHs and sends them to the BBU pool. Then, based on the obtained average channel gains, the BBU pool decides the cluster for each UE based on cluster formation techniques such as the ones in \cite{LiuTVT2016}. The dense C-RAN can be deployed in hot areas such as shopping mall and stadium, where the average channel gains change slowly and depends mainly on UEs' locations \cite{shenchen2016}. Hence, the clusters can be formed on a long time scale and are assumed to be fixed.

Denote ${\bf{h}}_{i,k} \in {{\mathbb{ C}}^{M \times 1}}$ and ${\bf{w}}_{i,k} \in {{\mathbb{ C}}^{M \times 1}}$ as the channel vector and the beam-vector from RRH $i$ to UE $k$, respectively. Then, the baseband received signal at UE $k$ is given by
\begin{equation}\label{receivedsignal}
  y_k = \underbrace {\sum\nolimits_{i \in {{\cal I}_k}} {{\bf{h}}_{i,k}^{\rm{H}}{\bf{w}}_{i,k}s_k} }_{{\rm{desired\  signal}}} + \underbrace {\sum\nolimits_{l \ne k,l \in {\cal U}} {\sum\nolimits_{i \in {{\cal I}_l}} {{\bf{h}}_{i,k}^{\rm{H}}{\bf{w}}_{i,l}s_l} } }_{ {\rm{interference}}} + z_k,
\end{equation}
where $s_k$ is the data symbol for UE $k$, $z_k$ is the additive complex white Gaussian noise that is assumed to follow the distribution  ${\cal C}{\cal N}(0,\sigma _k^2)$. It is assumed that ${{\mathbb{E}}}\{ |s_k{|^2}\}  = 1$ and ${{\mathbb{E}}}\{ s_{{k_1}}s_{{k_2}}\}  = 0$ for $k_1 \ne k_2, \forall k_1, k_2 \in {\cal U}$. The channel vector ${\bf{h}}_{i,k}$ can be decomposed as ${\bf{h}}_{i,k} = \sqrt {\alpha _{i,k}} {\bf{\bar h}}_{i,k}$, where $\alpha _{i,k}$ denotes the large-scale channel gain that includes the path loss and shadowing, and ${\bf{\bar h}}_{i,k}$ denotes the small-scale fading following the distribution of ${\cal C}{\cal N}({\bf{0}},{\bf{I}})$.

\vspace{-0.5cm}\subsection{Channel Estimation for Intra-cluster CSI}

To design the beam-vectors for all the UEs, the overall CSI over the network should be available at the BBU pool. However, it is an unaffordable task to acquire all CSI for dense C-RAN with large number of RRHs and UEs due to limited training resources. To handle this issue, we assume that each UE $k$ only measures the CSI to the RRHs in its cluster ${{\cal I}_k}$, while for the RRHs out of the cluster we assume that the BBU pool only knows the large-scale channel gains, i.e., $\{\alpha _{i,k}, \forall i\in {\cal I}\backslash {{\cal I}_k},k\in {\cal U}\}$. Note that the large-scale channel gains are used to manage the multiuser interference. If these gains are set to zero, the resulting solution may be too optimistic and may not satisfy UEs' quality of service (QoS) requirements as shown in \cite{Lakshmana2016}.

In this paper, we assume that $\tau$ time slots are used for channel training or equivalently the length of the pilot sequences is $\tau$, which is assumed to be smaller than the number of total UEs $K$ for the dense network, i.e., $\tau<K$. Note that the maximum number of orthogonal pilot sequences is equal to the pilot length $\tau$.  Hence, in orthogonal pilot assignment scheme, the maximum number of UEs that can be simultaneously served by C-RAN is upper bounded by $\tau$. To support more UEs, we propose a pilot reuse scheme in this paper.

Denote the available pilot set as ${\cal Q}=\{1,2,\cdots,\tau\}$, and the corresponding orthogonal pilot sequences as ${\bf{Q = }}\left[ {{{\bf{q}}_1}, \cdots ,{{\bf{q}}_\tau }} \right] \in {\mathbb{ C}^{\tau  \times \tau }}$, which satisfies the orthogonal condition, i.e.,  ${{\bf{Q}}^H}{\bf{Q}} = {\bf{I}}$. We denote an arbitrary pilot reuse scheme as ${\cal P}{({\cal U}, \cal Q)}=\{(k,\pi_k):k\in {{\cal U}}, \pi_k\in \cal Q\}$, where $(k,\pi_k)$ denotes that UE $k$ is allocated with pilot sequence  ${\bf{q}}_{\pi_k}$. In addition, define ${\cal K}_{\pi}=\{k:\pi_k=\pi\}$ as the set of UEs that reuse the $\pi$th pilot sequences.

Given the pilot reuse scheme ${\cal P}{( {\cal U}, \cal Q)}$, the UEs transmit their pilot sequences to the RRHs to estimate the channels during the uplink training phase. Specifically, the received pilot signal at RRH $i$ is
\vspace{-0.2cm}
\begin{equation}\label{equachaes}
  {{\bf{Y}}_i} = \sum\nolimits_{k \in{\cal U}} {\sqrt {{p_t}} {{\bf{h}}_{i,k}}{\bf{q}}_{\pi_k}^{\rm{H}}}  + {{\bf{N}}_i},
\end{equation}
where $p_t$ is the pilot transmit power at each UE, and ${{\bf{N}}_i} \in {{\mathbb C}^{M \times \tau }}$ is the additive Gaussian noise matrix received during the training phase, whose elements are independently generated and follow the distributions of ${\cal C}{\cal N}(0, \sigma^2)$.

To obtain channel ${{\bf{h}}_{i,k}},i \in {{\cal I}_k}$, the BBU pool first multiplies ${{\bf{Y}}_i}$ by ${{\bf{q}}_{\pi_k}}$, which yields
\begin{eqnarray}
  {{\bf{y}}_{i,k}} &=& \frac{1}{{\sqrt {{p_t}} }}{{\bf{Y}}_i}{{\bf{q}}_{\pi_k}} \nonumber\\
   &=& {{\bf{h}}_{i,k}} + {\sum _{l \in {{\cal K}_{{\pi _k}}}\backslash \{ k\} }}{{\bf{h}}_{i,l}}+ {{\bf{n}}_i}, \label{pioltcon}
\end{eqnarray}
where ${{\bf{n}}_i} = \frac{1}{{\sqrt {{p_t}} }}{{\bf{N}}_i}{{\bf{q}}_{\pi_k}}$. Since ${\bf{q}}_{\pi_k}$ is a unit-norm vector,  it is easy to show that ${{\bf{n}}_i}$ is still Gaussian distribution whose elements are independently and identically distributed as ${\cal C}{\cal N}(0, \frac{{{\sigma ^2}}}{{{p_t}}})$.  The MMSE estimate of channel ${{\bf{h}}_{i,k}} $ is given by
 \begin{equation}\label{detecton}
  {{{\bf{\hat h}}}_{i,k}} = \frac{{{\alpha _{i,k}}}}{{ \sum\nolimits_{l \in {{\cal K}_{\pi_k}}} {{\alpha _{i,l}}}  + {{\hat \sigma }^2}}}{{\bf{y}}_{i,k}}
 \end{equation}
with ${{\hat \sigma }^2} = {\sigma ^2}/p_t$. According to the property of MMSE estimate \cite{kailath2000linear}, channel estimation error ${{{\bf{\tilde h}}}_{i,k}} = {{\bf{h}}_{i,k}} - {{{\bf{\hat h}}}_{i,k}}$ is  independently distributed as ${\cal C}{\cal N}({\bf{0}},{\delta _{i,k}}{\bf{I}})$, where ${\delta _{i,k}}$ is given by
\begin{equation}\label{epl}
   {\delta _{i,k}} = \frac{{{\alpha _{i,k}}\left( {\sum\nolimits_{l \in {{\cal K}_{\pi_k}\backslash \{k\}}} {{\alpha _{i,l}}}  + {{\hat \sigma }^2}} \right)}}{{ \sum\nolimits_{l \in {{\cal K}_{\pi_k}}} {{\alpha _{i,l}}}  + {{\hat \sigma }^2}}}.
 \end{equation}

In the next two sections, we propose a two-stage optimization method to optimize the transmission in dense C-RAN. Specifically, in Stage I, we provide a joint pilot allocation and UE selection algorithm to maximize the number of admitted UEs with fixed number of available pilots in Section \ref{stage1}; in Stage II,  based on the results from Stage I, a robust beam-vector optimization algorithm is proposed to minimize the transmit power in Section \ref{userselectionalg}, while considering the pilot contamination.

\section{Stage I: Pilot Allocation and UE Selection Design}\label{stage1}

In this stage, we aim to design the pilot allocation scheme and UE selection method to maximize the number of admitted UEs under several practical constraints with fixed number of pilot sequences.

\vspace{-0.5cm}\subsection{Problem Formulation}

First, in conventional TDD communication systems, all UEs in the same macro-cell should be allocated with orthogonal training resources for the macro-BS to distinguish the channels from the UEs \cite{Biguesh2006}. Similarly, in dense C-RAN, the UEs served by the same RRH should be allocated with orthogonal training sequences. This kind of constraints can be mathematically expressed as follows:
\vspace{-0.1cm}
\begin{equation}\label{cons1}
 {\rm{C1:}}\ {\bf{q}}_{\pi_k}^{\rm{H}}{{\bf{q}}_{\pi_k'}} = 0, {\rm{for}}\  k,k' \in {{\cal U}_i},k \ne k',\forall i \in {\cal I}.\vspace{-0.1cm}
\end{equation}
For example, in Fig.~\ref{fig1}, RRH 11 jointly serves UE 1, UE 4 and UE 6. Then the following conditions should be satisfied: ${\bf{q}}_{\pi_1}^{\rm{H}}{{\bf{q}}_{\pi_4}} = 0, {\bf{q}}_{\pi_1}^{\rm{H}}{{\bf{q}}_{\pi_6}} = 0,$  and ${\bf{q}}_{\pi_4}^{\rm{H}}{{\bf{q}}_{\pi_6}} = 0$.

Second, to reduce the channel estimation error, the reuse times for each pilot sequence should be restricted under a predefined value. Denote the number of UEs that share pilot $l$ as $n_l$, then the constraints can be expressed as
\vspace{-0.2cm}
\begin{equation}\label{reusetime}
 {\rm{C2:}}\  n_l\le n_{\rm{max}}, \forall l \in \cal Q.\vspace{-0.2cm}
\end{equation}
where $n_{\rm{max}}$ is the maximum reuse time for each pilot. Constraint C2 means that we should reuse the pilot sequences in a fair way to avoid the extreme case where one pilot sequence is reused many times while there are several unused pilot sequences. This extreme case will deteriorate the system performance.

Here, we aim to find the maximum number of UEs that can be served with a fixed number of pilot sequences. Hence, the joint pilot allocation and UE selection problem at Stage I can be formulated as
\vspace{-0.2cm}
\begin{subequations}\label{stageI}
\begin{align}
{\cal P}_1:\ \mathop {\max }\limits_{{{ {\cal U}}}\subseteq {\bar {\cal U}}, {\cal P}{( {\cal U}, \cal Q)}} \quad
& \left| \cal U \right|
\\
\textrm{s.t.}\quad {\rm{ C1,C2}}.  \nonumber \vspace{-0.5cm}
\end{align}
\end{subequations}

Problem ${\cal P}_1$ can be considered as a resource-assignment problem, which is difficult to solve in general. The optimal solution can be obtained by the exhaustive search method, where one should check all possible pilot allocation schemes and choose one with the maximum number of admitted UEs while satisfying the constraints C1 and C2. However, the complexity of the exhaustive search method increases exponentially with the number of UEs, which is not practical for dense C-RAN. In the following, we provide a low-complexity pilot allocation scheme, which is suitable for practical applications.

\vspace{-0.5cm}\subsection{Pilot Allocation Scheme}
\begin{figure}
\centering
\includegraphics[width=4.5in]{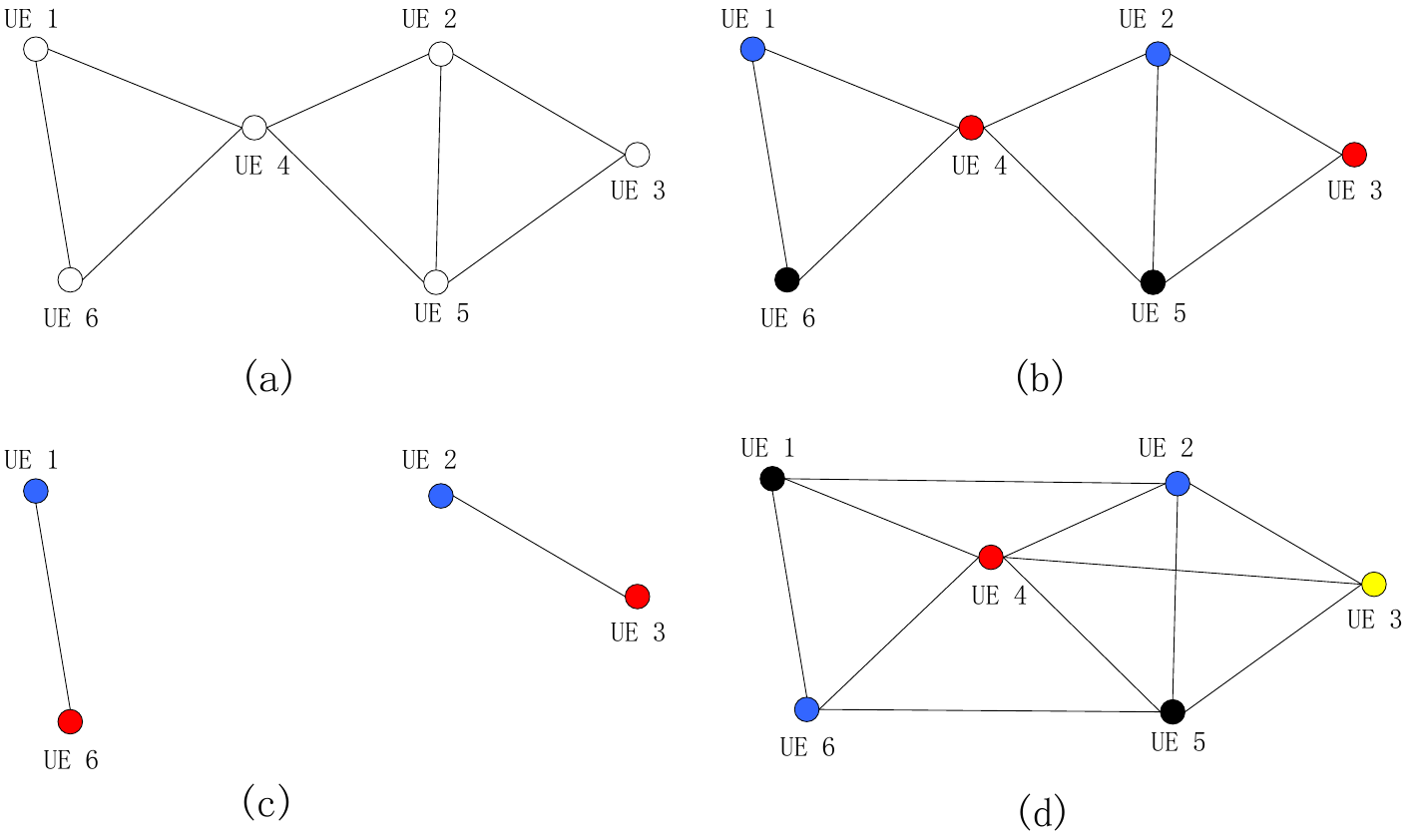}\vspace{-0.5cm}
\caption{(a) Construction of the undirected graph for the network in Fig.~\ref{fig1}, where the vertexes denotes the UEs and any two UEs sharing at least one RRH should be connected with each other; (b) The colored graph after applying the Dsatur algorithm \cite{Chen2016tvt} with $n_{\rm{max}}=2$, the minimum number of pilot sequences required is $n^*=3$; (c) The UE selection and pilot reallocation result after applying Algorithm \ref{algorithmcase1} when $\tau=2$ and $n_{\rm{max}}=2$, the number of selected UEs is four; (d) The pilot reallocation result after applying Algorithm \ref{algorithmcaseII} when $\tau=4$ and $n_{\rm{max}}=2$. In this colored
graph, UE 1 and UE 2 use different pilots to avoid the pilot interference, and the same holds for UE 5 and UE6. In addition, UE 3 and UE 4 are allocated with different pilots, which additionally reduces the pilot interference. }\vspace{-0.8cm}
\label{fig2}
\end{figure}

Constraints C1 can be represented by a $K \times K$ binary matrix ${\bf{B}}$, where the $k$th row and ${k'}$th column  of matrix ${\bf{B}}$ is given by
\vspace{-0.2cm}
\begin{equation}\label{conventionalB}
  {b}_{k,k'} = \left\{ \begin{array}{l}
1,\ {\rm{   if  }}\ {{\cal I}_k} \cap {{\cal I}_{k'}} \ne \emptyset \ {\rm{ and  }}\ k \ne k'\\
0,\ {\rm{  otherwise.}}
\end{array} \right.
\end{equation}
When any two UEs are served by at least one common RRH, the corresponding element in $\bf{B}$ should be one and these two UEs should be allocated with different pilots. Otherwise, the element is zero and the UEs can share the same pilot. Based on the matrix ${\bf{B}}$, we can construct an undirected graph to describe the relationship between any two UEs for constraint C1, where any two UEs that are served by at least one common RRH should be connected with each other. For the network in Fig.~\ref{fig1},  the corresponding graph is shown in Fig.~\ref{fig2} (a), where any two UEs that are served by at least one common RRH are connected together. The graph coloring algorithm such as the Dsatur algorithm \cite{brelaz1979new}, which aims for coloring the vertexes of a graph with the minimum number of different colors under the same constraints in C1 and C2 for a given set of UEs,  has been used in  \cite{Chen2016tvt} for user-centric C-RAN to design the pilot allocation.  For the undirected graph in Fig.~\ref{fig2} (a), the final colored graph is shown in Fig.~\ref{fig2} (b) after using the Dsatur algorithm. In this example, the minimum number of colors (pilots) required is three.  The computational complexity of this algorithm is low.  However, given the set of pilot sequences, how to design the pilot allocation scheme jointly with the UE selection process that maximizes the number of admitted UEs  needs further investigation.
%\vspace{-0.5cm}
%\begin{algorithm}
%\caption{Dsatur algorithm }\label{conalgorithmDsa}
%\begin{algorithmic}[1]
%\STATE Initialize the matrix $\bf{B}$, the UE set ${\cal U} =\bar{\cal U}=\left\{ {1, \cdots ,K} \right\}$, initial pilot set $\tilde {\cal Q}{\rm{ = }}\emptyset $, total number of different pilots $n=0$;
%\STATE While ${\cal U}  \ne \emptyset $
%\STATE \qquad Find ${k^*} = \mathop {\arg \max }\nolimits_{k \in {\cal U}} {\theta_k}$. If there exists more than one UE with the same ${\theta_k}$ values, select one  \\ \qquad  with the most number of unallocated neighbors; \\
%\STATE \qquad Assign pilot to UE ${k^*}$.  \\
% \qquad 1)\ If there are available pilots in $\tilde {\cal Q}$, find ${l^*} = \mathop {\arg \min }\nolimits_{l \in \widetilde {\cal{Q}}} {n_l}$, allocate pilot ${l^*}$ to UE $k^*$, i.e.,  \\
% \qquad\quad   $\pi_{k^*}=l^*$. Increase the reuse number of this pilot by one, $n_{l^*}=n_{l^*}+1$.  If this number is  \\
% \qquad\quad equal to $n_{\rm{max}}$, remove it from $\tilde {\cal Q}$, i.e., $\widetilde {\cal Q} = \widetilde {\cal Q}\backslash \{ {l^*}\} $;  \\
% \qquad 2)\ If not, increase $n$ by one, i.e., $n=n+1$, and allocate pilot $n$ to UE ${k^*}$, i.e.,  $\pi_{k^*}=n$.  Add  \\
% \qquad\quad  the new pilot into the pilot set $\tilde {\cal Q}$, i.e., $\widetilde {\cal Q} = \widetilde {\cal Q} \cup \{ n\} $.
% \STATE \qquad Remove UE $k^*$ from ${\cal U} $;
%\end{algorithmic}
%\end{algorithm}
%\vspace{-0.5cm}
To resolve this issue, we first adopt the Dsatur algorithm to find the minimum number of colors that are required for all the UEs in $\bar {\cal U}$ while satisfying constraints C1 and C2. If the minimum number of required pilots is larger than the number of available pilots, some UEs should be removed. Otherwise, all UEs can be serve. For the latter case, although all users can be admitted, the pilot allocation results may be that some pilots have not been allocated by the Dsatur algorithm while some pilots are reused by up to $n_{\rm{max}}$ users, which wastes the pilot resources. In this case, we can reallocate all of the available pilots to the UEs to reduce the pilot contamination. The details of each case will be discussed in the following.

Denote the minimum number of pilots required by the Dsatur algorithm as $n^*$. In the following, we discuss two cases: 1) $n^*> \tau$; 2) $n^*< \tau$. When $n^*=\tau$, no additional operation is required.

%Specifically, we first define a metric $\eta_{k,k'}$ to measure the level of the pilot contamination between any two unconnected UEs when they are allocated with the same pilot,
%\begin{equation}\label{meause}
%  {\eta _{k,k'}} = \log \left( {1 + \frac{{\sum\nolimits_{i \in {{\cal I}_{k'}}} {{\alpha _{i,k}}} }}{{\sum\nolimits_{i \in {{\cal I}_k}} {{\alpha _{i,k}}} }}} \right) + \log \left( {1 + \frac{{\sum\nolimits_{i \in {{\cal I}_k}} {{\alpha _{i,k'}}} }}{{\sum\nolimits_{i \in {{\cal I}_{k'}}} {{\alpha _{i,k'}}} }}} \right).
%\end{equation}
%Obviously, larger ${\eta _{k,k'}}$ means more severe pilot contamination between UE $k$ and UE $k'$ when the same pilot is allocated to them. Then, define ${\xi _{\pi_k}} = \sum\nolimits_{k' \in {{\cal K}_{\pi_k}}} {{\eta _{k,k'}}} $ as the value to measure the level of the introduced pilot contamination when pilot $\pi_k$ is allocated to UE $k$. Fig.\ref{fig2}-(a) and (b) show the colored results after applying the conventional and improved Dsatur algorithms to the undirected graph constructed from the network in Fig.\ref{fig1}, respectively. In Fig.\ref{fig2}-(a), UE 1 and UE 2 are not so far away from each other, but they reuse the same pilot, which may have measurable pilot interference. The same issue holds for UE 5 and UE 6. However, from Fig.\ref{fig2}-(b), it is seen that UE 1 and UE 2 have been allocated with different pilots, which alleviates the pilot interference.

\emph{Case I: $n^*>\tau$}. In this case, some UEs should be removed so that the minimum number of required pilots is no larger than $\tau$. Denote ${\theta_k}\buildrel \Delta \over = \sum\nolimits_{k' \ne k,k' \in \bar{\cal U}} {{b_{k,k'}}}$ as the total number of different UEs to which UE $k$ is connected to. In general, the UE with the largest $\theta_k$ should be removed since many UEs should use different pilots from that used by this UE, which will increase the number of required pilots. However, there are some cases that different UEs share the same largest $\theta_k$ value, and randomly removing one UE may incur inferior performance. Intuitively, the UE that incurs the highest pilot contamination should be removed. To this end, we first define a metric $\eta_{k,k'}$ to measure the level of  pilot contamination between any two unconnected UEs when they are allocated with the same pilot,
\begin{equation}\label{meause}
  {\eta _{k,k'}} = \log \left( {1 + \frac{{\sum\nolimits_{i \in {{\cal I}_{k'}}} {{\alpha _{i,k}}} }}{{\sum\nolimits_{i \in {{\cal I}_k}} {{\alpha _{i,k}}} }}} \right) + \log \left( {1 + \frac{{\sum\nolimits_{i \in {{\cal I}_k}} {{\alpha _{i,k'}}} }}{{\sum\nolimits_{i \in {{\cal I}_{k'}}} {{\alpha _{i,k'}}} }}} \right).
\end{equation}
The above definition of ${\eta _{k,k'}}$ is inspired by the channel estimation error in (\ref{epl}). Obviously, the larger ${\eta _{k,k'}}$ means more severe pilot contamination between UE $k$ and UE $k'$ when the same pilot is allocated to them. Then, define ${\xi _k} = \sum\nolimits_{k' \in {\cal K}_{\pi_k}\backslash \{k\}} {{\eta _{k,k'}}} $ as the value to measure the level of pilot contamination when keeping UE $k$. Then, the UE with the largest value of ${\xi _k}$ should be removed.

Based on this idea, we provide a UE selection and pilot reallocation process in Algorithm \ref{algorithmcase1}. By using this algorithm for the case of $\tau=2$ and $n_{\rm{max}}=2$ for the network in Fig.~\ref{fig1}, the UE selection and pilot allocation result is shown in Fig.~\ref{fig2} c), where the number of selected UEs is four.

\vspace{-0.5cm}
\begin{algorithm}
\caption{UE selection and pilot reallocation algorithm for Case I}\label{algorithmcase1}
\begin{algorithmic}[1]
\STATE Initialize the matrix $\bf{B}$, the UE set ${\cal U} =\bar{\cal U}=\left\{ {1, \cdots ,K} \right\}$, the initial number of required pilots  $n^*$ obtained from the Dsatur algorithm;
\STATE While $n^*  > \tau $
\STATE \qquad Find ${k^*} = \mathop {\arg \max }\nolimits_{k \in {\cal U}} {\theta_k}$. If there are more than one UE with the same values of $\theta_k$,  \\
\qquad  remove  the UE with the largest ${\xi _k} $;
\STATE \qquad Remove UE $k^*$ from ${\cal U}$, i.e., ${\cal U}{\rm{ = }}{\cal U}/{k^*}$, and update matrix $\bf{B}$ with current ${\cal U}$;
\STATE \qquad Use the  Dsatur algorithm to calculate  $n^*$ with $\bf{B}$ and ${\cal U}$;
\end{algorithmic}
\end{algorithm}
\vspace{-0.5cm}

\emph{Case II: $n^*<\tau$.} In this case, all UEs can be admitted in this stage and only part of the available pilots are allocated. However, all the available pilots should be allocated to UEs to reduce the pilot contamination as much as possible. For example, in Fig.~\ref{fig2} b) with three allocated pilots, there may exist measurable pilot interference between UE 1 and UE 2 since they are not so far away from each other. A similar issue holds for UE 3 and UE 4 or UE 5 and UE 6. When there are four available pilots, the pilots can be fully used and reallocated to resolve this issue. For example, in Fig.~\ref{fig2} d), when there are four pilots, the UEs in each pair are able to be allocated with different pilots. Hence, the pilot contamination can be additionally mitigated.

As the definition of ${\eta _{k,k'}}$ in (\ref{meause}) shows, larger ${\eta _{k,k'}}$ means more severe pilot contamination between UE $k$ and UE $k'$ when the same pilot is allocated to them. The pair of UEs with larger ${\eta _{k,k'}}$ should be allocated with different pilots, or equivalently these two UEs should be connected with each other. For the pair of UEs with smaller ${\eta _{k,k'}}$, they can be allocated  the same pilot. Hence, to reconstruct the undirected graph, a threshold $\eta_{\rm{th}}$ is introduced to determine whether two UEs can reuse the same pilot. Specifically, when ${\eta _{k,k'}}>\eta_{\rm{th}}$, no reuse is allowed for these two UEs. Otherwise, these two UEs can reuse the same pilot. Based on $\eta_{\rm{th}}$, the binary matrix $\bf{B}$ can be reconstructed as follows
\begin{equation}\label{reconsB}
 {b}_{k,k'} = \left\{ \begin{array}{l}
1,\ {\rm{   if  }}\ {{\cal I}_k} \cap {{\cal I}_{k'}} \ne \emptyset \ {\rm{ and  }}\ k \ne k',\\
1,\ {\rm{if}}\  {\eta _{k,k'}} > {\eta _{{\rm{th }}}},\ {{\cal I}_k} \cap {{\cal I}_{k'}} = \emptyset\ {\rm{ and  }}\ k \ne k', \\
0,\ {\rm{  otherwise.}}
\end{array} \right.
\end{equation}
Obviously, when $\eta_{\rm{th}}$ is small, more UEs will be connected with each other and more orthogonal pilots are required. In the extreme case when $\eta_{\rm{th}}<{\rm{min}}\{{\eta}_{k,k'}\}$, all UEs will be connected with each other and the number of required pilots is equal to the number of total UEs $K$. On the other hand, when $\eta_{\rm{th}}$ is very large, less number of UEs will be connected and the number of required pilots will decrease. In the extreme case when $\eta_{\rm{th}}\geq {\rm{max}}\{{\eta}_{k,k'}\}$, the reconstructed binary matrix $\bf{B}$ in (\ref{reconsB}) reduces to the conventional binary matrix $\bf{B}$ in (\ref{conventionalB}), where  the pairs of UEs without sharing the same RRHs will not be connected with each other. The number of required pilots in this case will be equal to  $n^*$. As it is assumed that $\tau<K$, there must exist at least one $\eta_{\rm{th}}$ between
${\rm{min}}\{{\eta}_{k,k'}\}$ and ${\rm{max}}\{{\eta}_{k,k'}\}$ that the number of required pilots is equal to $\tau$. As a result, the bisection search method can be adopted to find the $\eta_{\rm{th}}$ such that the required number of pilots is equal to $\tau$. The details are given in Algorithm \ref{algorithmcaseII}. Fig.~\ref{fig2} d) shows the pilot allocation results after using Algorithm \ref{algorithmcaseII}.
\vspace{-0.5cm}
\begin{algorithm}
\caption{Pilot reallocation algorithm for Case II}\label{algorithmcaseII}
\begin{algorithmic}[1]
\STATE Initialize the lower-bound $\eta_{\rm{th},\rm{LB}}={\rm{min}}\{{\eta}_{k,k'}\}$, the upper-bound $\eta_{\rm{th},\rm{UB}}={\rm{max}}\{{\eta}_{k,k'}\}$, the initial number of required pilots  $n^*$ obtained from the Dsatur algorithm;
\STATE While $n^*  \ne  \tau $
\STATE \qquad Set ${\eta _{{\rm{th}}}} = {{\left( {{\eta _{{\rm{th}},{\rm{LB}}}} + {\eta _{{\rm{th}},{\rm{UB}}}}} \right)} \mathord{\left/
 {\vphantom {{\left( {{\eta _{{\rm{th}},{\rm{LB}}}} + {\eta _{{\rm{th}},{\rm{UB}}}}} \right)} 2}} \right.
 \kern-\nulldelimiterspace} 2}$, update the binary matrix $\bf{B}$ in (\ref{reconsB}). Use the  Dsatur \\
 \qquad     algorithm to  calculate  $n^*$ with $\bf{B}$.
\STATE \qquad If $n^*>\tau$, set $\eta_{\rm{th},\rm{LB}}={\eta _{{\rm{th}}}}$; If $n^*<\tau$, set $\eta_{\rm{th},\rm{UB}}={\eta _{{\rm{th}}}}$;
\end{algorithmic}
\end{algorithm}

\vspace{-0.5cm}\subsection{Complexity analysis}
Now, we provide the complexity analysis for the proposed pilot assignment and UE selection scheme. The computational complexity of the Dsatur algorithm in \cite{Chen2016tvt} is on the order of ${\cal O}\left( {{K^2}} \right)$.

For Case I, the user selection algorithm in Algorithm \ref{algorithmcase1} requires the Dsatur algorithm  to be executed for at most $K$ times to find the final set of admitted UEs, and thus the total complexity is on the order of ${\cal O}\left( {{K^3}} \right)$. For Case II, when $\eta _{{\rm{th}}}$ falls in a specific region $[a,c]$, the number of required pilots  $n^*$ obtained from the  Dsatur algorithm  will be equal to $\tau$. Hence, when $\eta_{\rm{th},\rm{UB}}-\eta_{\rm{th},\rm{LB}}<c-a$, Algorithm \ref{algorithmcaseII} will terminate. As a result, the maximum number of iterations required by Algorithm \ref{algorithmcaseII} is upper bounded by ${\log _2}\left( {{{\left| {{\rm{max}}\{ {\eta_{k,k'}}\}  - {\rm{min}}\{ {\eta_{k,k'}}\} } \right|} \mathord{\left/
 {\vphantom {{\left| {{\rm{max}}\{ {\eta_{k,k'}}\}  - {\rm{min}}\{ {\eta_{k,k'}}\} } \right|} {(c - a)}}} \right.
 \kern-\nulldelimiterspace} {(c - a)}}} \right)$. As a result, the total complexity of Algorithm \ref{algorithmcaseII} is ${\cal O}\left( {{K^2}{{\log }_2}\left( {{{\left| {{\rm{max}}\{ {\eta_{k,k'}}\}  - {\rm{min}}\{ {\eta_{k,k'}}\} } \right|} \mathord{\left/
 {\vphantom {{\left| {{\rm{max}}\{ {\eta_{k,k'}}\}  - {\rm{min}}\{ {\eta_{k,k'}}\} } \right|} {(c - a)}}} \right.
 \kern-\nulldelimiterspace} {(c - a)}}} \right)} \right)$.

For comparison, we also provide the complexity analysis for the exhaustive search method. For a given set of UEs $\cal U$ with $\left| {\cal U} \right| = l$, the total number of pilot allocation results is given by ${\tau ^l}$. For each pilot allocation, one should check whether constraints C1 and C2 are satisfied or not. The total number of possible UE sets is given by  $\sum\nolimits_{l = 1}^K {C_K^l}$. As the result, the total complexity of the exhaustive search method is  $\sum\nolimits_{l = 1}^K {C_K^l{\tau ^l}}$, which is much higher than the proposed algorithm and will be unaffordable for dense C-RAN with large number of UEs.

\section{Stage II: Robust Beamforming Design}\label{userselectionalg}
In this stage, we first formulate the beamforming optimization problem to minimize the transmit power with the UEs selected from Stage I while considering the effect of pilot contamination due to pilot reuse scheme in Section \ref{stage1}. Then, a low-complexity beam-vector optimization algorithm is proposed to solve this optimization problem along with the complexity analysis.

\vspace{-0.6cm}\subsection{Problem formulation}

Denote the set of UEs selected from Stage I as $\widetilde  {\cal U}$. The beam-vector for each UE can be merged into a single large-dimension vector  $ { {{\bf{ w}}}}_k= {[ {{\bf{w}}_{i,k}^{{\rm{H}}},\forall i \in {{\cal I}_k}} ]^{\rm{H}}} \in {{\mathbb{C}}^{\left| {{{\cal I}_k}} \right|M \times 1}}$. Similarly, we define a set of new channel vectors  ${\bf{ g}}_{l,k} = [ {{\bf{h}}_{i,k}^{\rm{H}},\forall i \in {{\cal I}_l}} ]^{\rm{H}} \in {{\mathbb C}^{\left| {{{\cal I}_l}} \right|M \times 1 }}$, representing the aggregated perfect CSI from the RRHs in ${\cal I}_l$ to UE $k$. Also, define ${\bf{\tilde g}}_{k,k}= [{\bf{\tilde h}}_{i,k}^{\rm{H}},\forall i \in {{\cal I}_k}]^{\rm{H}}  \in {{\mathbb C}^{\left| {{{\cal I}_k}} \right|M \times 1 }}$ and ${\bf{ \hat g}}_{k,k} = [{\bf{\hat h}}_{i,k}^{\rm{H}},\forall i \in {{\cal I}_k}]^{\rm{H}}\in {{\mathbb C}^{\left| {{{\cal I}_k}} \right|M \times 1}}$ as the aggregated CSI error and estimated CSI from the RRHs in ${\cal I}_k$ to UE $k$, respectively. Since channel estimation error is expressed as ${{{\bf{\tilde g}}}_{k,k}} = {{\bf{g}}_{k,k}} - {{{\bf{\hat g}}}_{k,k}}$, the received signal model in (\ref{receivedsignal}) can be rewritten as
\vspace{-0.2cm}
\begin{equation}\label{achieveablerate}
  {y_k} = \underbrace {{\bf{\hat g}}_{k,k}^{\rm{H}}{{\bf{w}}_k}{s_k}}_{{\rm{Desired\  signal}}} + \underbrace {{\bf{\tilde g}}_{k,k}^{\rm{H}}{{\bf{w}}_k}{s_k}}_{{\rm{Self - interference}}} + \underbrace {\sum\nolimits_{l \ne k,l \in \widetilde{\cal U}} {{\bf{g}}_{l,k}^{\rm{H}}{{\bf{w}}_l}{s_l}} }_{{\rm{ Interference\  from\  other\  UEs}}} + {z_k},\forall k \in \widetilde {\cal U}.
\end{equation}
As in most of existing works \cite{Jose2011,wence2015}, we consider the achievable data rate where the term corresponding to the channel estimation error in (\ref{achieveablerate}) is regarded as Gaussian noise. Specifically, the achievable data rate for UE $k$ is written as
\vspace{-0.1cm}
\begin{equation}\label{datarateUEk}
  {r_k} = \frac{{T - \tau }}{T}{\mathbb{E}}\left\{ {{{\log }_2}\left( {1 + \frac{{{{\left| {{\bf{\hat g}}_{k,k}^{\rm{H}}{{\bf{w}}_{k}}} \right|}^2}}}{{{{\left| {{\bf{\tilde g}}_{k,k}^{\rm{H}}{{\bf{w}}_k}} \right|}^2} + \sum\nolimits_{l \ne k,l \in \widetilde{\cal U}} {{{\left| {{\bf{g}}_{l,k}^{\rm{H}}{{\bf{w}}_l}} \right|}^2} + \sigma _k^2} }}} \right)} \right\}, \forall k\in \widetilde {\cal U},
\end{equation}
where $T$ denotes the coherence time of the channel in terms of time slots, the expectation is taken with respect to the unknown channel estimation errors $\left\{ {{{{\bf{\tilde h}}}_{i,k}},i \in {{\cal I}_k},\forall k \in \widetilde {\cal U}} \right\}$, and the small-scale inter-cluster CSI $\left\{ {{{\bf{h}}_{i,k}},i \in {\cal I}\backslash {{\cal I}_k}} \right\}$.
Each UE's data rate should be larger than its minimum data rate requirement ${R_{k,\min }}$:
\vspace{-0.3cm}
\begin{equation}\label{ratecons}
  {\rm{C3:}}\  {r}_{k}  \ge {R_{k,\min }},\forall k \in \widetilde {\cal U}.\vspace{-0.2cm}
\end{equation}

Since the capacity of fronthaul links is upper bounded by the available bandwidth, each fronthaul link should be imposed with the fronthaul capacity constraints:
\vspace{-0.2cm}
\begin{equation}\label{fronthaulca}
  {\rm{C4}:}\ \sum\nolimits_{k \in {\cal U}_i} {\varepsilon \left({\left\| {{\bf{w}}_{i,k}} \right\|}^2\right){r}_{k}}  \le {C_{i,\max }},\forall i \in {\cal I},\vspace{-0.2cm}
\end{equation}
where  $\varepsilon \left( \cdot \right)$ is an indicator function,  defined as $\varepsilon \left( x \right) = 1$ if $x\ne 0$, otherwise, $\varepsilon \left( x \right) = 0$, and $C_{i,\max }$ is the maximum data rate that can be supported by the $i$th fronthaul link.

Finally, each RRH has its own power constraint, represented as
\begin{equation}\label{powerconst}
{\rm{C5}}:\;\sum\nolimits_{k \in {{\cal U}_i}} {{\left\| {{\bf{w}}_{i,k}} \right\|}^2}  \le {P_{i,\max }},i \in {\cal I},
\end{equation}
where ${P_{i,\max }}$ is the power constraint of RRH $i$.

In Stage II, we aim to jointly optimize the UE-RRH associations and beam-vectors to minimize the total transmit power of the dense C-RAN network, while guaranteeing the rate constraints in C3, the fronthaul capacity constraints in C4, and the per-RRH power constraints in C5. Specifically, this optimization problem can be formulated as
\begin{equation}\label{transmitpower}
\begin{array}{l}
 {\cal P}_2:\ \mathop {\min }\limits_{{\bf{w}} } \quad \sum\nolimits_{k \in {\widetilde{\cal U}}} {\sum\nolimits_{i \in {\cal I}} {{{\left\| {{{\bf{w}}_{i,k}}} \right\|}^2}} } \\
\qquad\ {\rm{s}}.{\rm{t}}.\qquad {\kern 1pt} {\rm{C3}},{\rm{C4}},{\rm{C5}},
\end{array}
\end{equation}
where $\bf{w}$ denotes the collection of all beam-vectors.

The imperfect intra-cluster CSI and incomplete inter-cluster CSI make the accurate closed-form expression of the data rate difficult to obtain. In the following, we first obtain the lower-bound of the data rate and replace it with this term to make the optimization problem in (\ref{transmitpower}) more tractable. By using Jensen's inequality, the lower bound of the data rate can be derived as
\begin{eqnarray}
 r_k &\ge& \frac{{T - \tau }}{T}{\log _2}\left( {1 + \frac{{{{\left| {{\bf{\hat g}}_{k,k}^{\rm{H}}{{\bf{w}}_{k}}} \right|}^2}}}{{{\mathbb{E}}\left\{ {{{\left| {{\bf{\tilde g}}_{k,k}^{\rm{H}}{{\bf{w}}_k}} \right|}^2}} \right\} + \sum\nolimits_{l \ne k,l \in \widetilde{\cal U}} {\mathbb{E}}{\left\{ {{{\left| {{\bf{g}}_{l,k}^{\rm{H}}{{\bf{w}}_l}} \right|}^2}} \right\} + \sigma _k^2} }}} \right)\label{firstineq}\\
 &=& \frac{{T - \tau }}{T}{\log _2}\left( {1 + \frac{{{{\left| {{\bf{\hat g}}_{k,k}^{\rm{H}}{{\bf{w}}_{k}}} \right|}^2}}}{{{\bf{w}}_k^{\rm{H}}{{\bf{E}}_{k,k}}{{\bf{w}}_k} + \sum\nolimits_{l \ne k,l \in \widetilde{\cal U}} {{\bf{w}}_l^{\rm{H}}{{\bf{A}}_{l,k}}{{\bf{w}}_l} + \sigma _k^2} }}} \right)\label{secondg}\\
 &\buildrel \Delta \over =&  \tilde r_k,\label{thirdp}
 \end{eqnarray}
where ${{\bf{E}}_{k,k}} = {\rm{blkdiag}}\left\{ {{\delta _{i,k}}{{\bf{I}}_{M \times M}},i \in {{\cal I}_k}} \right\}$, and ${\bf{A}}_{l,k} = {{\mathbb{E}}}\left\{ {{\bf{g}}{{_{l,k}^{\rm{H}}}}{\bf{g}}_{l,k}} \right\}\in {\mathbb{C}}^{M{|{\cal I}_l|} \times M{|{\cal I}_l|}}$. To obtain the expression of ${\bf{A}}_{l,k} $, we define the indices of ${{{\cal I}_l}}$ as ${{\cal I}_l} = \{ s_1^l, \cdots ,s_{|{{\cal I}_l}|}^l\} $. Then, we have
\begin{equation}\label{amatrix}
  {{\bf{A}}_{l,k}} = \left[ {\begin{array}{*{20}{c}}
{{{\left( {{{\bf{A}}_{l,k}}} \right)}_{1,1}}}& \cdots &{{{\left( {{{\bf{A}}_{l,k}}} \right)}_{1,|{{\cal I}_l}|}}}\\
 \vdots & \ddots & \vdots \\
{{{\left( {{{\bf{A}}_{l,k}}} \right)}_{|{{\cal I}_l}|,1}}}& \cdots &{{{\left( {{{\bf{A}}_{l,k}}} \right)}_{|{{\cal I}_l}|,|{{\cal I}_l}|}}}
\end{array}} \right],l \ne k,
\end{equation}
where ${\left( {{\bf{A}}_{l,k}} \right)_{i,j}} \in {{\mathbb{C}}^{M \times M}},i,j \in 1, \cdots ,|{{\cal I}_l}|$  is the block matrix of ${\bf{A}}_{l,k}$ at the $i$th row and $j$th column,   given by
\begin{equation}\label{ak}
 {\left( {{{\bf{A}}_{l,k}}} \right)_{i,j}} = \left\{ \begin{array}{l}
{{{\bf{\hat h}}}_{s_i^l,k}}{\bf{\hat h}}_{s_j^l,k}^{\rm{H}},\;\;\;\;\;\qquad \qquad \ \ {\rm{if}}\;s_i^l,s_j^l \in {{\cal I}_k},i \ne j,\\
{{{\bf{\hat h}}}_{s_i^l,k}}{\bf{\hat h}}_{s_j^l,k}^{\rm{H}} + {\delta  _{s_i^l,k}}{{\bf{I}}_{M \times M}},\ \ {\rm{if}}\;s_i^l,s_j^l \in {{\cal I}_k},i = j,\\
{{\alpha _{s_i^l,k}}{{\bf{I}}_{M \times M}},\;\qquad \qquad\quad\  {\rm{if}}\;s_i^l,s_j^l \notin {{\cal I}_k},{\rm{and}}\;i = j,}\\
{{\bf{0}}_{M \times M,}}\qquad\ \  \ \ \qquad\qquad\ {\rm{otherwise.}}
\end{array} \right.
\end{equation}
It can be easily verified that ${\bf{A}}_{l,k}$ is a positive definite matrix.  In \cite{pan2017jsac}, we provided the accurate data rate expression for the special case of non-overlapped cluster. We showed that the approximation error was  within 3\% in the considered setups.

By replacing the data rate $r_k$ in data rate constraint ${\rm{C3}}$ with its lower-bound or achievable data rate $\tilde r_k$ given in (\ref{thirdp}), constraint ${\rm{C3}}$ can be rewritten as
\begin{equation}\label{constrc}
{\rm{C6:}} \  {\tilde r}_{k}  \ge {R_{k,\min }},\forall k \in \widetilde {\cal U},
\end{equation}
which can be equivalently rewritten as
\begin{equation}\label{equa}
{\rm{C7}}:\frac{{{{\left| {{\bf{\hat g}}_{k,k}^{\rm{H}}{{\bf{w}}_k}} \right|}^2}}}{{{\bf{w}}_k^{\rm{H}}{{\bf{E}}_{k,k}}{{\bf{w}}_k} + \sum\nolimits_{l \ne k,l \in \cal U} {{\bf{w}}_l^{\rm{H}}{{\bf{A}}_{l,k}}{{\bf{w}}_l} + \sigma _k^2} }} \ge {\eta _{k,\min }}
\end{equation}
where ${\eta _{k,\min }} = {2^{\frac{{{R_{k,\min }}T}}{{T - \tau }}}} - 1$. In addition, by replacing the data rate $r_k$ in fronthaul constraints ${\rm{C4}}$ with its lower-bound $\tilde r_k$, constraints ${\rm{C4}}$ can be rewritten as
\begin{equation}\label{frontulca}
  {\rm{C8}:}\ \sum\nolimits_{k \in {\cal U}_i} {\varepsilon \left({\left\| {{\bf{w}}_{i,k}} \right\|}^2\right){\tilde r}_{k}}  \le {C_{i,\max }},\forall i \in {\cal I}.
\end{equation}
Then, Problem ${\cal P}_2$ can be approximately transformed as
\begin{subequations}\label{appstaone}
\begin{align}
{\cal P}_3:\ \mathop {\min }\limits_{{\bf{w}} } \quad
& \sum\nolimits_{k \in {\widetilde{\cal U}}} {\sum\nolimits_{i \in {\cal I}} {{{\left\| {{{\bf{w}}_{i,k}}} \right\|}^2}} }
\\
\qquad\ \textrm{s.t.}\qquad\!\!\!\!
&{\rm{C5}}, {\rm{C7}}, {\rm{C8}}.
\end{align}
\end{subequations}
To simplify the fronthaul capacity constraints C8, we first provide the following theorem.

\itshape \textbf{Theorem 1:}  \upshape Denote the optimal solution of Problem ${\cal P}_3$ as ${{\bf{w}}^ \star } = \left[ {{\bf{w}}_j^ \star {\rm{,}}j \in \widetilde {\cal U}} \right]$. Then the minimum rate constraints are met with equality at the optimal point, i.e., ${{\tilde r}_k}({{\bf{w}}^ \star }) = {R_{k,\min }},\forall k \in \widetilde {\cal U}$.

\itshape \textbf{Proof:}  \upshape Please see Appendix \ref{rateequlity}. \hfill $\Box$

Based on Theorem 1, the expressions of rate  lower bound ${\tilde r}_{k}$'s in constraint C8 can be replaced by their rate targets, which yields the following optimization problem
\begin{subequations}\label{appstaone}
\begin{align}
{\cal P}_4:\ \mathop {\min }\limits_{{\bf{w}} } \quad
& \sum\nolimits_{k \in {\widetilde{\cal U}}} {\sum\nolimits_{i \in {\cal I}} {{{\left\| {{{\bf{w}}_{i,k}}} \right\|}^2}} }
\\
\qquad\ \textrm{s.t.}\qquad\!\!\!\!
&{\rm{C5}}, {\rm{C7}}, \\
&{\rm{C9:}}\ \sum\nolimits_{k \in {\cal U}_i} {\varepsilon \left( {\left\| {{\bf{w}}_{i,k}} \right\|}^2\right){R_{k,\min }}}  \le {C_{i,\max }},\forall i \in {\cal I}.
\end{align}
\end{subequations}

However, Problem ${\cal P}_4$ is still difficult to solve due to the following reasons. First, due to the per-RRH power constraints of C5 and the fronthaul capacity constraints of C9, the system may not be able to support all the UEs that are selected in Stage I. Second, constraint C9 contains the non-smooth and non-differential indicator function, which is usually named as an mixed-integer non-linear programming (MINLP) problem that is NP-hard to solve. One can solve the problem via the exhaustive search method. Specifically, for each given set of UE-RRH associations, one should check whether Problem ${\cal P}_4$ is feasible or not, if feasible, then the problem can be solved.  Hence, the complexity is on the order of $O\left( {{2^{ \left|\widetilde {\cal U} \right|I}}} \right)$, which is prohibitive for dense C-RAN with large number of UEs and RRHs. Third, due to the presence of the channel estimation error, the conventional weighted minimum mean square error (WMMSE) method \cite{Qingjiang2011} that has been used in \cite{pan2017joint,pan2017jsac,Binbin2014} for the perfect intra-cluster CSI case cannot be used for the problem considered here. Hence, even given the set of selected UEs and UE-RRH associations, it is still difficult to check the feasibility of Problem ${\cal P}_4$.

\vspace{-0.5cm}\subsection{Low-complexity Algorithm}

\subsubsection{UE selection method}
Denote the set of admitted UEs as $\cal U$. Inspired by the UE selection algorithm in \cite{Matskani2008}, we construct the following alternative optimization problem by introducing a series of auxiliary non-negative variables $\{x_k\}_{k\in {{\cal U}}} $:
\vspace{-0.2cm}
\begin{subequations}\label{UEsel}
\begin{align}
{\cal P}_5:\ &\mathop {\mathop {\min }\limits_{{\bf{w}},{{\{ {x_k}\geq0\} }_{k \in{\cal U}}}} } \
 \sum\nolimits_{i \in {{\cal I}}} {\sum\nolimits_{k \in {{\cal U}_i}} {{{\left\| {{{\bf{w}}_{i,k}}} \right\|}^2}} } +\Gamma{\sum\nolimits_{k \in{\cal U}} {x_k} }
\\
\ \textrm{s.t.}\quad\!\!\!\!
&{\rm{C5}}, {\rm{C9}},\\
&{\rm{C10}}:\;{{\left| {{\bf{\hat g}}_{k,k}^{\rm{H}}{{\bf{w}}_k}} \right|}^2} + {x_k}  \ge {\eta _{k,\min }}\left( {{\bf{w}}_k^{\rm{H}}{{\bf{E}}_{k,k}}{{\bf{w}}_k} + \sum\nolimits_{l \ne k,l \in {\cal U}} {{\bf{w}}_l^{\rm{H}}{{\bf{A}}_{l,k}}{{\bf{w}}_l} + \sigma _k^2} } \right),\forall k
\end{align}
\end{subequations}
where $\Gamma$ is a large constant. Obviously, Problem ${\cal P}_5$ is always feasible.  If the optimal solution $\{{x_k},\forall k\in  {\cal U} \}$ are equal to zeros, i.e., $\{{x_k}=0,\forall k\in {\cal U} \}$,  all UEs can be admitted. Otherwise, some UEs should be removed. The large constant $\Gamma$ acts as a penalty factor that forces as many $x_k$s to zero as possible,  so that the number of admitted users is maximized. Hence, the objective function of Problem ${\cal P}_5$ is to maximize the number of admitted users while simultaneously minimizing the transmit power.

Intuitively, the UE with the largest $x_k$ should have the highest priority to be removed since it has the largest gap from the rate target. Based on this idea, we provide  a low-complexity algorithm to solve Problem ${\cal P}_5$ in Algorithm \ref{selctalg}. The main idea is to remove the UE with the largest $x_k$ in each iteration.

\vspace{-0.5cm}
\begin{algorithm}
\caption{Low-complexity UE selection Algorithm}\label{selctalg}
\begin{algorithmic}[1]
\STATE  Initialize the set of users ${\cal U}=\widetilde{\cal U}$ from Stage I;
\STATE  Given ${\cal U}$,  solve Problem ${\cal P}_5$ to obtain $\left\{ {x _k} \right\}_{k \in {\cal U}}$ and ${{\bf{w}}}$;
 \STATE If $x_k=0, \forall k\in {\cal U}$,  output ${\bf{w}}$ and ${\cal U}^{\star}\!\!=\!\!{\cal U}$, terminate; Otherwise, find  ${k^{\star}} \!\!=\!\! \arg \mathop {\max }\nolimits_{k\in {\cal U}} x_k$, remove user $k^\star$ and update ${\cal U} ={\cal U}\backslash {k^{\star}}$, go to step 2.
\end{algorithmic}
\end{algorithm}
\vspace{-0.5cm}
\subsubsection{Method to deal with the indicator function}
Similar to \cite{pan2017jsac}, we approximate the non-smooth indicator function as a fractional function ${f_\theta }(y) = \frac{y}{{y + \theta }}$, where $\theta$ is a small positive value. By replacing the indicator function in Problem ${\cal P}_5$ with ${f_\theta }(y)$, Problem ${\cal P}_5$ can be approximated as
\vspace{-0.2cm}
\begin{subequations}\label{appro}
\begin{align}
{\cal P}_6:\ \mathop {\mathop {\min }\limits_{{\bf{w}},{{\{ {x_k}\geq 0\} }_{k \in {\cal U}}}} } \quad
& \sum\nolimits_{k \in {{\cal U}}} {\sum\nolimits_{i \in {\cal I}} {{{\left\| {{{\bf{w}}_{i,k}}} \right\|}^2}} } +\Gamma{\sum\nolimits_{k \in {\cal U}} {x _k} }
\\
\qquad\ \textrm{s.t.}\qquad\!\!\!\!
&{\rm{C5}}, {\rm{C10}},{\rm{C11}}:\sum\nolimits_{k \in {{\cal U}_i}} {{f_\theta }\left( {{{\left\| {{{\bf{w}}_{i,k}}} \right\|}^2}} \right){R_{k,\min }}}  \le {C_{i,\max }},\forall i \in {\cal I}.
\end{align}
\end{subequations}
Although constraints $\rm{C11}$ are still non-convex since ${f_\theta }(x)$ is a concave function, it is the  difference of convex (d.c.) program,
which can be efficiently solved by the successive convex approximation (SCA) method \cite{dinh2010local}. The main idea is to approximate it as its first-order Taylor expansion. Specifically, by using the concavity of ${f_\theta }(x)$, we have
\vspace{-0.2cm}
\begin{equation}\label{fucntions}
 {f_\theta }\left( {{{\left\| {{{\bf{w}}_{i,k}}} \right\|}^2}} \right) \le {f_\theta }\left( {{{\left\| {{{\bf{w}}_{i,k}}(t)} \right\|}^2}} \right) + { \beta }_{i,k}(t) \left( { {{{\left\| {{{\bf{w}}_{i,k}}} \right\|}^2}}  - {{{\left\| {{{\bf{w}}_{i,k}}(t)} \right\|}^2}} } \right)\vspace{-0.2cm}
\end{equation}
where ${\bf{w}}_{i,k}(t)$ is the  beam-vector obtained at the $t^{\rm{th}}$ iteration, $\beta_{i,k} (t) = {f_\theta ^\prime }\left( {{{\left\| {{{\bf{w}}_{i,k}}(t)} \right\|}^2}} \right)$\footnote{$f_{\theta }^\prime (x)$ denotes the first-order derivative of $f_{\theta }(x)$ w.r.t. $x$.}. By replacing ${f_\theta }\left( {{{\left\| {{{\bf{w}}_{i,k}}} \right\|}^2}} \right)$ in Problem ${\cal P}_6$ with the right hand side of (\ref{fucntions}),  the optimization problem to be solved in the $(t+1)^{\rm{th}}$ iteration is given by
\vspace{-0.2cm}
\begin{subequations}\label{eachprob}
\begin{align}
{\cal P}_7:\ {\mathop {\min }\limits_{{\bf{w}},{{\{ {x_k\geq 0}\} }_{k \in {\cal U}}}} } \quad
& \sum\nolimits_{k \in {{\cal U}}} {\sum\nolimits_{i \in {\cal I}} {{{\left\| {{{\bf{w}}_{i,k}}} \right\|}^2}} } +\Gamma{\sum\nolimits_{k \in {\cal U}} {x_k} }
\\
\qquad\ \textrm{s.t.}\qquad\!\!\!\!
&{\rm{C5}}, {\rm{C10}},{\rm{C12}}:\sum\nolimits_{k \in {{\cal U}_i}} {{\tau _{i,k}}(t){{\left\| {{{\bf{w}}_{i,k}}} \right\|}^2}}  \le {{\tilde C}_{i }}(t), \forall i \in {\cal I},
\end{align}
\end{subequations}
where ${\tau _{i,k}}(t) = {\beta _{i,k}}(t){R_{k,\min }}$, ${{\tilde C}_i}(t) = {C_{i,\max }} - \sum\nolimits_{k \in {{\cal U}_i}} {\left( {{f_\theta }\left( {{{\left\| {{{\bf{w}}_{i,k}}(t)} \right\|}^2}} \right) - {\beta _{i,k}}(t){{\left\| {{{\bf{w}}_{i,k}}(t)} \right\|}^2}} \right){R_{k,\min }}} $.

Based on the above analysis, we provide the SCA  algorithm to solve Problem ${\cal P}_6$ in Algorithm \ref{algorithmitersca}. According to \cite{dinh2010local,Pan2016jsac,pan2017jsac},  the convergence of the SCA algorithm can be guaranteed under two conditions: 1) The initial beam-vectors are feasible for Problem ${\cal P}_6$; 2) In each iteration of the SCA algorithm, the globally optimal solution of Problem ${\cal P}_7$ can be obtained. The first condition can be easily guaranteed, while the second condition is in general difficult to satisfy due to the non-convex constraints of C10. Fortunately, we are able to obtain the globally optimal solution of Problem ${\cal P}_7$ as shown in the following subsection.
\vspace{-0.5cm}
\begin{algorithm}
\caption{SCA Algorithm to Solve Problem ${\cal P}_6$}\label{algorithmitersca}
\begin{algorithmic}[1]
\STATE Initialize  the iteration number $t=1$,  error tolerance $\delta $. Initialize any feasible ${\bf{w}}{(0)}$, calculate ${\tau _{i,k}}(0)$, ${{\tilde C}_i}(0)$,  calculate the objective value of Problem ${\cal P}_6$, denoted as ${\rm{Obj(}}{{\bf{w}}{(0)}}{\rm{)}}$.
 \STATE Solve Problem ${\cal P}_7$ to get ${\bf{w}}{(t)}$ with ${\tau _{i,k}}(t-1)$, ${{\tilde C}_i}(t-1)$;
 \STATE Update ${\tau _{i,k}}(t)$, ${{\tilde C}_i}(t)$ with ${\bf{w}}{(t)}$;
 \STATE If  ${{\left| {{\rm{Obj(}}{{\bf{w}}{(t - 1)}}{\rm{) - Obj(}}{{\bf{w}}{(t)}}{\rm{)}}} \right|} \mathord{\left/
 {\vphantom {{\left| {{\rm{Obj(}}{{\bf{V}}^{(n - 1)}}{\rm{) - Obj(}}{{\bf{V}}^{(n)}}{\rm{)}}} \right|} {{\rm{Obj(}}{{\bf{w}}^{(t)}}{\rm{)}}}}} \right.
 \kern-\nulldelimiterspace} {{\rm{Obj(}}{{\bf{w}}{(t)}}{\rm{)}}}} < \delta  $, terminate.  Otherwise, set $t \leftarrow t + 1$, go to step 2.
\end{algorithmic}
\end{algorithm}
\vspace{-0.5cm}

\subsubsection{Method to deal with the non-convex constraint C10}

We apply the semi-definite relaxation approach \cite{Gershman2010} to solve Problem ${\cal P}_7$. Specifically,
define ${{\bf{W}}_k} = {{\bf{w}}_k}{\bf{w}}_k^{\rm{H}},\forall k\in {\cal U}$ with the constraints that ${\rm{rank}}({\bf{W}}_k)=1,\forall k \in {\cal U}$. Then, Problem ${\cal P}_7$ can be equivalently transformed as\footnote{For simplicity, the iteration index $t$ is omitted. }
\begin{subequations}\label{sdpprob}
\begin{align}
{\cal P}_8:\ \mathop {\mathop {\min }\limits_{{{\{{\bf{W}}_k\succeq {\bf{0}}, {x _k\geq0}\} }_{k \in {\cal U}}}} } \quad
& {\sum\nolimits_{k \in {\cal U}} {({\rm{tr}}\left( {{{\bf{W}}_k}} \right)+\Gamma x _k)} }
\\
\qquad\ \textrm{s.t.}\qquad\!\!\!\!
&{\rm{C}}13: \sum\nolimits_{k \in {{\cal U}_i}} {{\rm{tr}}\left( {{{\bf{W}}_k}{{\bf{B}}_{i,k}}} \right)}  \le {P_{i,\max }},\forall i \in {\cal I},\\
&
{\rm{C}}14: {\rm{tr}}\left( {{{\bf{W}}_k}{{{\bf{\hat g}}}_{k,k}}{\bf{\hat g}}_{k,k}^{\rm{H}}} \right) + {x_k}\nonumber\\
 &\ge {\eta _{k,\min }}\left( {{\rm{tr}}\left( {{{\bf{W}}_k}{{\bf{E}}_{k,k}}} \right) + \sum\nolimits_{l \ne k,l \in {\cal U}} {{\rm{tr}}\left( {{{\bf{W}}_l}{{\bf{A}}_{l,k}}} \right) + \sigma _k^2} } \right),\forall k \in {\cal U},\\
 &{\rm{C}}15: \sum\nolimits_{k \in {{\cal U}_i}} {{\tau _{i,k}}{{\rm{tr}}\left( {{{\bf{W}}_k}{{\bf{B}}_{i,k}}} \right)}}  \le {{\tilde C}_{i }}, \forall i \in {\cal I},\\
 &{\rm{C}}16: {\rm{rank(}}{{\bf{W}}_k}{\rm{) = 1}},\forall k \in {\cal U},
\end{align}
\end{subequations}
where  ${{\bf{B}}_{i,k}}$'s are the following block diagonal matrices
\begin{equation}\label{blocmat}
  {{\bf{B}}_{i,k}}{\rm{ = diag}}\left\{ {\overbrace {{{\bf{0}}_{1 \times M}}}^{s_1^k}, \cdots ,\overbrace {{{\bf{1}}_{1 \times M}}}^{s_j^k},\overbrace {{{\bf{0}}_{1 \times M}}}^{s_{j + 1}^k}, \cdots ,\overbrace {{{\bf{0}}_{1 \times M}}}^{s_{\left| {{{\cal I}_{ k}}} \right|}^k}} \right\},\ {\rm{if }}\ s_j^k = i,\forall i \in {\cal I},k \in {\cal U}.
\end{equation}
However, Problem ${\cal P}_8$ is still non-convex due to the rank-one constraints in C16. We further relax these non-convex constraints and obtain the following optimization  problem:
\begin{subequations}\label{sdpprobconv}
\begin{align}
{\cal P}_9:\ \mathop {\mathop {\min }\limits_{{{\{{\bf{W}}_k\succeq {\bf{0}}, {\alpha _k}\} }_{k \in {\cal U}}}} } \quad
& {\sum\nolimits_{k \in{\cal U}} {({\rm{tr}}\left( {{{\bf{W}}_k}} \right)+\alpha _k^2)} }
\\
\qquad\ \textrm{s.t.}\qquad\!\!\!\!
&{\rm{C}}13, {\rm{C}}14, {\rm{C}}15.
\end{align}
\end{subequations}
Obviously, Problem ${\cal P}_9$ is a semi-definite programming (SDP) problem \cite{boyd2004convex}, which is convex and can be effectively solved by using the standard tools such as CVX.

In general, the optimal solution obtained from SDP Problem may not satisfy the rank-one constraints. Fortunately, for Problem ${\cal P}_9$, we can prove that the relaxation of the non-convex constraints is tight. Denote the optimal solution of ${\cal P}_9$ as ${\bf{W}}_k^{\star}, \alpha _k^{\star}, \forall k$, we have the following theorem.

\itshape \textbf{Theorem 2:}  \upshape  The optimal solution obtained from the SDP Problem ${\cal P}_9$ is guaranteed to satisfy the rank-one constraints, i.e., ${\rm{rank}}\left( {{\bf{W}}_k^ \star } \right) = 1,\forall k$.

\itshape \textbf{Proof:}  \upshape Please see Appendix \ref{prooftheorem1}. \hfill $\Box$

Since the rank of ${\bf{W}}_k^{\star}, \forall k$ are equal to one, one can use the simple singular value decomposition operation to obtain the optimal beam-vector of Problem ${\cal P}_7$. Hence, the SCA algorithm will be guaranteed to converge.

\vspace{-0.5cm}\subsection{Complexity analysis}

For simplicity, we assume the candidate size for each UE is equal to $L$, i.e., $|{\cal I}_k|=L, \forall k \in \cal U$. The method to solve Problem ${\cal P}_4$ consists of two layers. The inner layer is to solve the SDP problem ${\cal P}_9$ by using the interior-point method. According to \cite{lobo1998applications}, the number of iterations required to reduce the duality gap to a constant fraction is upper-bounded by $O\left( {\sqrt {(\left| {\cal U} \right|+2I)ML} } \right)$. The total number of real variables in Problem ${\cal P}_4$ is given by $N_{\rm{tot}}={\left| {\cal U} \right|\left( {\frac{{ML\left( {ML{\rm{ + }}1} \right)}}{2}{\rm{ + }}1} \right)}$. Then, the complexity for each iteration of the interior-point method is given by $O\left( {N_{{\rm{tot}}}^2\left| {\cal U} \right|{M^2}{L^2}} \right)$ \cite{lobo1998applications}. Hence, the complexity of the interior-point method to solve ${\cal P}_9$ is given by $O\left( {\sqrt {(\left| {\cal U} \right|+2L)ML} N_{{\rm{tot}}}^2\left| {\cal U} \right|{M^2}{L^2}} \right)$. According to the UE selection algorithm in Algorithm \ref{selctalg}, each UE is removed for each time, and thus Problem ${\cal P}_9$ needs to be solved at most ${\left| {\cal U} \right|}$ times in the worst case. Hence, the total complexity to solve Problem ${\cal P}_4$ is on the order of $O\left( {\sqrt {(\left| {\cal U} \right|+2L)ML} N_{{\rm{tot}}}^2{{\left| {\cal U} \right|}^2}{M^2}{L^2}} \right)$, which can be solved within polynomial time.

\vspace{-0.2cm}\section{Simulation Results}\label{simlresult}

In this section, simulation results are provided to evaluate the performance of the proposed algorithms. Two types of dense C-RAN networks are considered: small dense C-RAN deployed in a square area of 400 m $\times$ 400 m and large one with 700 m $\times$ 700 m. The numbers of UEs in small and large dense C-RANs are given by 8 and 24 with the densities of 50 UEs/km$^2$ and 49 UEs/km$^2$, respectively. The numbers of RRHs in both dense C-RANs are set as 14 and 40 with the densities of 87.5 RRHs/km$^2$ and 81.6 RRHs/km$^2$, respectively. The considered scenarios comply with the 5G ultra-dense network \cite{xiaohuge2016}, where the density of 5G BS is highly anticipated to come up to 40-50 BS/km$^2$. Both the UEs and RRHs are assumed to be independently and uniformly placed in this area. The channel gains are composed of three parts: 1) the channel path loss is modeled as $PL = 148.1 + 37.6{\log _{10}}d\ ({\rm{dB}})$ \cite{access2010further}, where  $d$ is the distance measured in km; 2) the log-normal shadowing fading with zero mean and 8 dB standard derivation; 3) Rayleigh fading with zero mean and unit variance.
%All UEs are assumed to have the same rate requirements, i.e., ${R_{\min }} = {R_{k,\min }}, \forall k$, and all fronthaul links have the same fronthaul capacity constraints, i.e., ${C_{\max }} = {C_{i,\max }},\forall i$. For ease of exposition, the normalized fronthaul capacity is considered, i.e., ${{\tilde C}_{\max }} = {{{C_{\max }}} \mathord{\left/
% {\vphantom {{{C_{\max }}} {{R_{\min }}}}} \right.
% \kern-\nulldelimiterspace} {{R_{\min }}}}$, which denotes the number of UEs that each fronthaul link can support.
It is assumed that each UE chooses its nearest $L$ RRHs as its serving cluster, i.e., $\left| {{{\cal I}_k}} \right| = L,\forall k$. Unless otherwise specified, the other system parameters are set as follows: the number of transmit antennas at each RRH $M=2$,  system bandwidth $B=20\  \rm{MHz}$, error tolerance $\delta=10^{-5}$,  noise power spectral density is -174 dBm/Hz, each RRH's maximum power $P_{i,\rm{max}}=100\  {\rm{mW}}, \forall i$, large constant $\Gamma=10^5$, the pilot power at each UE is $p_t=200 \  {\rm{mW}}$, the parameter $\theta$ in the fractional function is $\theta=10^{-5}$, the minimum rate requirement for each UE and the capacity constraint for each fronthaul link are uniformly generated within the regions of $\left[ {3,5} \right]$ bit/s/Hz and $\left[ {5,10} \right]$ bit/s/Hz, respectively, the cluster size for each UE $L = 3$, the proportion of pilots for training in one coherence time is 1$\%$ \cite{Stands}, The pilot maximum reuse times  for small dense C-RAN and large one are set as  $n_{\rm{max}}=2$ and $n_{\rm{max}}=4$, respectively. The MMSE channel estimation method is used in our simulations. The following results are obtained by averaging over 200 independent trials, where in each trial, both the UEs and RRHs are randomly placed.

We compare our proposed algorithms (with legend ``Proposed") with the following algorithms:
\begin{enumerate}
  \item Orthogonal pilot allocation (with legend ``Ortho"): The number of admitted UEs in Stage I is equal to the number of available pilots $\tau$, and $\tau$ UEs are randomly selected from $K$ UEs.
  \item No reallocation operations for Case II in Stage I (with legend ``NoCaseII''): This approach is similar to the proposed pilot allocation method in Stage I, except that when Case II happens, no additional operation is applied to re-allocate the pilots.
  \item Conventional pilot allocation method (with legend ``Con''): This approach is similar to the above approach, except that when Case I happens, UEs are randomly removed until the minimum number of pilots is equal to $\tau$.
   \item Perfect CSI estimation (with legend ``Perfect''): In this approach, we assume that the CSI within each UE's serving cluster can be perfectly known.  Note that it is not necessary to consider the pilot allocation stage, and  the set of UEs selected from Stage I of the proposed algorithm is used as the initial set of UEs in Stage II of this approach.
  \item Exhaustive search method (with legend ``Exhau''): In this approach, when Case I in Stage I happens, exhaustive search method is adopted to find the maximum number of UEs that can be admitted. If Case II happens, exhaustive search method is used to find all pilot allocation results that the number of allocated pilots is equal to $\tau$. In Stage II, exhaustive search method is used to find the maximum number of admitted UEs. Note that the exhaustive search method has an exponential complexity, which is only feasible for small networks.
\end{enumerate}
\vspace{-0.5cm}\subsection{Small C-RAN networks}

\begin{figure}
\begin{minipage}[t]{0.475\linewidth}
\centering
\includegraphics[width=2.7in]{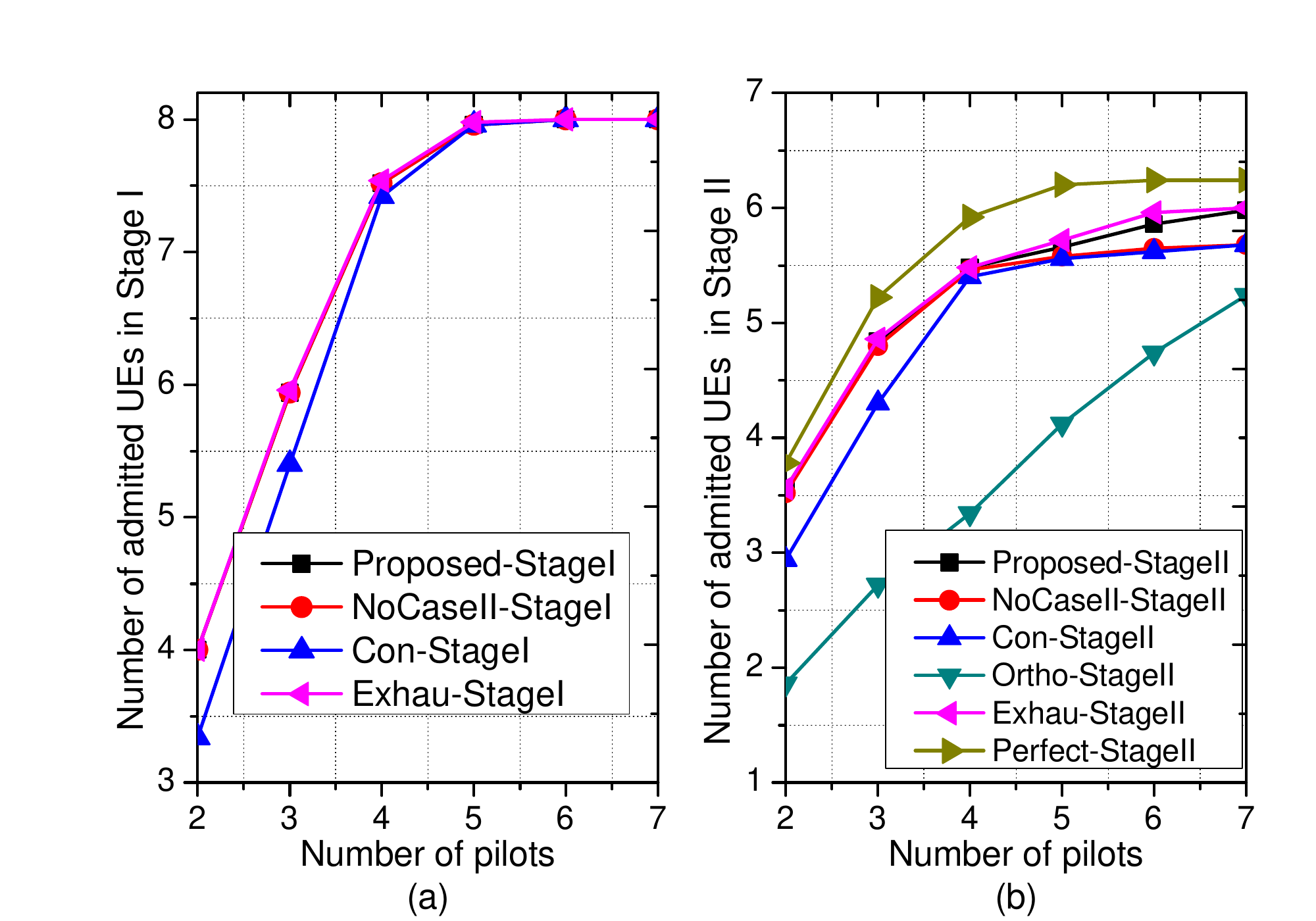}\vspace{-0.4cm}
\caption{(a) Number of admitted UEs in Stage I versus the number of available pilots $\tau$ for small networks. (b) Number of admitted UEs in Stage II versus the number of available pilots $\tau$ for small networks. }\vspace{-0.5cm}
\label{fig3}\vspace{-0.4cm}
\end{minipage}%
\hfill
\begin{minipage}[t]{0.475\linewidth}
\centering
\includegraphics[width=2.5in]{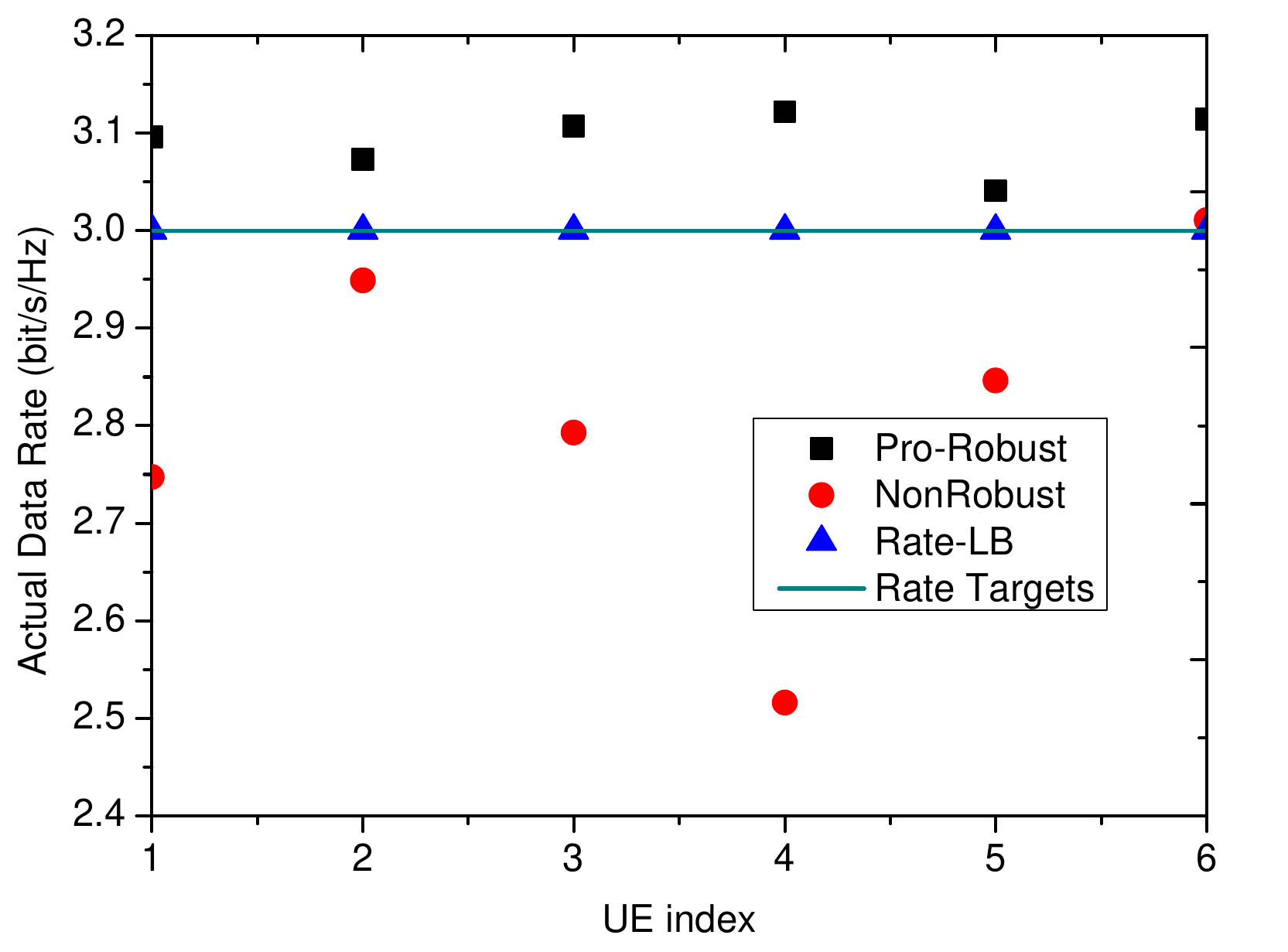}\vspace{-0.4cm}
\caption{The actual data rate for various algorithms under small dense C-RAN networks when six UEs are finally admitted in Stage II.}\vspace{-0.5cm}
\label{fig4}\vspace{-0.4cm}
\end{minipage}%
\hfill
\end{figure}

We first consider the small C-RAN networks  for the facilitation of applying the exhaustive search method. Fig.~\ref{fig3} shows the number of admitted UEs versus the number of available pilots $\tau$ for Stage I and Stage II, while Fig.~\ref{fig4} verifies the robustness of our proposed algorithm.

From Fig.~\ref{fig3}-(a), we can observe that the numbers of admitted UEs for all algorithms increase with the number of available pilots.  When  $\tau\leq 4$, our proposed algorithm outperforms the ``Con'' algorithm. The reason is that the ``Con'' algorithm randomly removes the UE without considering the pilot interference among the UEs. With the increase of $\tau$, the performance gain shrinks and finally both algorithms have the same performance where all UEs can be admitted by using the Dsatur algorithm. It is observed from Fig.~\ref{fig3}-(a) that the proposed algorithm and the ``NoCaseII'' have almost the same performance over the entire range of $\tau$. This is not surprising, as when Case I in Stage I occurs, both the proposed algorithm and the ``NoCaseII'' algorithm employ Algorithm \ref{algorithmcase1} to remove the UEs. On the other hand, when Case II happens, both algorithms can support all the UEs. The superiority of the proposed algorithm over the ``NoCaseII'' algorithm will be observed in Fig.~\ref{fig3}-(b) discussed later. Note that the exhaustive search method performs almost the same as the proposed algorithm. However, the former one requires enormous computational complexity, which is not feasible for large-scale networks. The performance of the ``Ortho'' algorithm is not shown in this figure. It has the worst performance since no UEs are allowed to reuse the pilots, and the number of admitted UEs is equal to the number of available pilots.

It is also seen from Fig.~\ref{fig3}-(b) that the numbers of admitted UEs in Stage II for all the algorithms increase with the number of available pilots. The performance of the proposed algorithm  is comparable to that of the exhaustive search method over the entire range of $\tau$, but the former needs huge computational complexity. The proposed algorithm outperforms the
``NoCaseII'' algorithm when $\tau\geq 4$, and the performance gain increases with $\tau$. This can be explained as follows. When $\tau$ increases, Case II is more likely to happen, our proposed algorithm reallocates the unassigned pilots to the UEs to additionally reduce the pilot interference, which leads to more accurate CSI. Thus, more UEs can be admitted. On the other hand, no reallocation operations are considered in ``NoCaseII'', which will incur severe pilot interference and less UEs can be admitted. This highlights the necessity of reallocating the unassigned pilots to UEs.   It is also observed from Fig.~\ref{fig3}-(b) that as $\tau$ increases, our proposed algorithm approaches the performance of the ``Perfect'' algorithm, which verifies the effectiveness of the proposed algorithm.

Now, we study the robustness of our proposed algorithm in Fig.~\ref{fig4} for one randomly generated simulation, where the actual data rates achieved by various algorithms are plotted. For the clarity of illustration, the data rate targets and fronthaul capacity constraints are set to be the same for all UEs and fronthaul links, which are given by ${R_{\min }} = 3$ bit/s/Hz and ${C_{\min }} = 9$ bit/s/Hz, respectively.  For comparison, the performance of non-robust algorithm is also shown (labeled as `Nonrobust'), where channel estimation error is not taken into account and the estimated channel is naively regarded as the perfect channel.  The data rate lower-bound obtained in (\ref{thirdp}) is also shown in this figure which is labeled as `Rate-LB'. The bold line in Fig.~\ref{fig4} represents the rate targets of all UEs. It is observed that the rate lower bound is equal to their data rate target, which verifies the correctness of Theorem 1. As expected, the actual data rates achieved by all UEs are above the bold line, which means that our proposed algorithm can provide guaranteed data rates for all UEs. Note that the gap between the rate lower bound and actual data rates achieved by our proposed algorithm are within $0.1$ bit/s/Hz, which is acceptable in practice. In contrast, the actual data rates achieved by the non-robust algorithm for almost all UEs are below the bold line, and the maximum gap is up to 0.5 bit/s/Hz for UE 4.

%From Fig.~\ref{fig5},  we can observe that the total transmit power for all the algorithms increases with $\tau$ when $\tau\leq 4$. This is due to the fact that the number of admitted UEs increase rapidly as seen in Fig.~\ref{fig4}, and more power is needed to support these UEs. However, when $\tau>4$, the total power for almost all algorithms fluctuates. This uncertainty may be attributed to the following two reasons: First, when
%$\tau>4$, the number of admitted UEs increases slowly which leads to the slow increased power; Second, when $\tau$ increases, the channel estimation error reduces and CSI will become more accurate that leads to the reduced power. It is difficult to determine which one will counteract the other one, which in turn incurs the fluctuation of the transmit power.

\vspace{-0.5cm}\subsection{Large-scale C-RAN networks}

In this subsection, we study the effects of different system parameters on the performance of the algorithms in large-scale dense C-RAN networks.
% In the following examples, we can also observe the fluctuations of the total transmit power for almost all algorithms. In addition, according to the problem formulations in Problem ${\cal P}_4$, the highest priority is to maximize the number of admitted UEs. Hence, the performance of total transmit power does not provide meaningful insights for analysis and will not be shown in the following.

\subsubsection{Impacts of candidate size}
\begin{figure}
\begin{minipage}[t]{0.475\linewidth}
\centering
\includegraphics[width=3.0in]{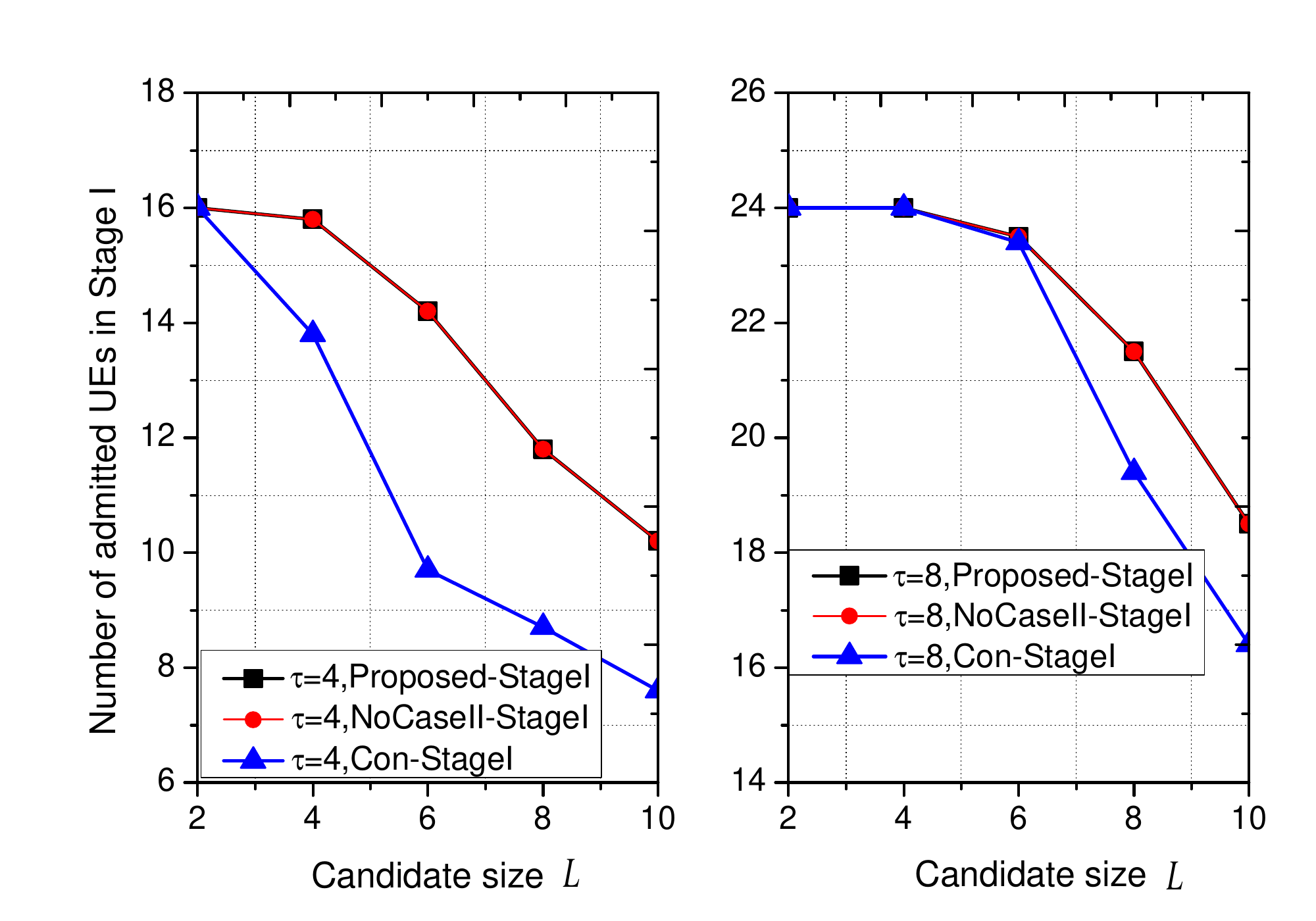}\vspace{-0.4cm}
\caption{Number of admitted UEs in Stage I versus candidate size $L$. The left subplot corresponds to the case of $\tau=4$ while the right one is $\tau=8$.}\vspace{-0.8cm}
\label{fig6}\vspace{-0.4cm}
\end{minipage}%
\hfill
\begin{minipage}[t]{0.475\linewidth}
\centering
\includegraphics[width=3.0in]{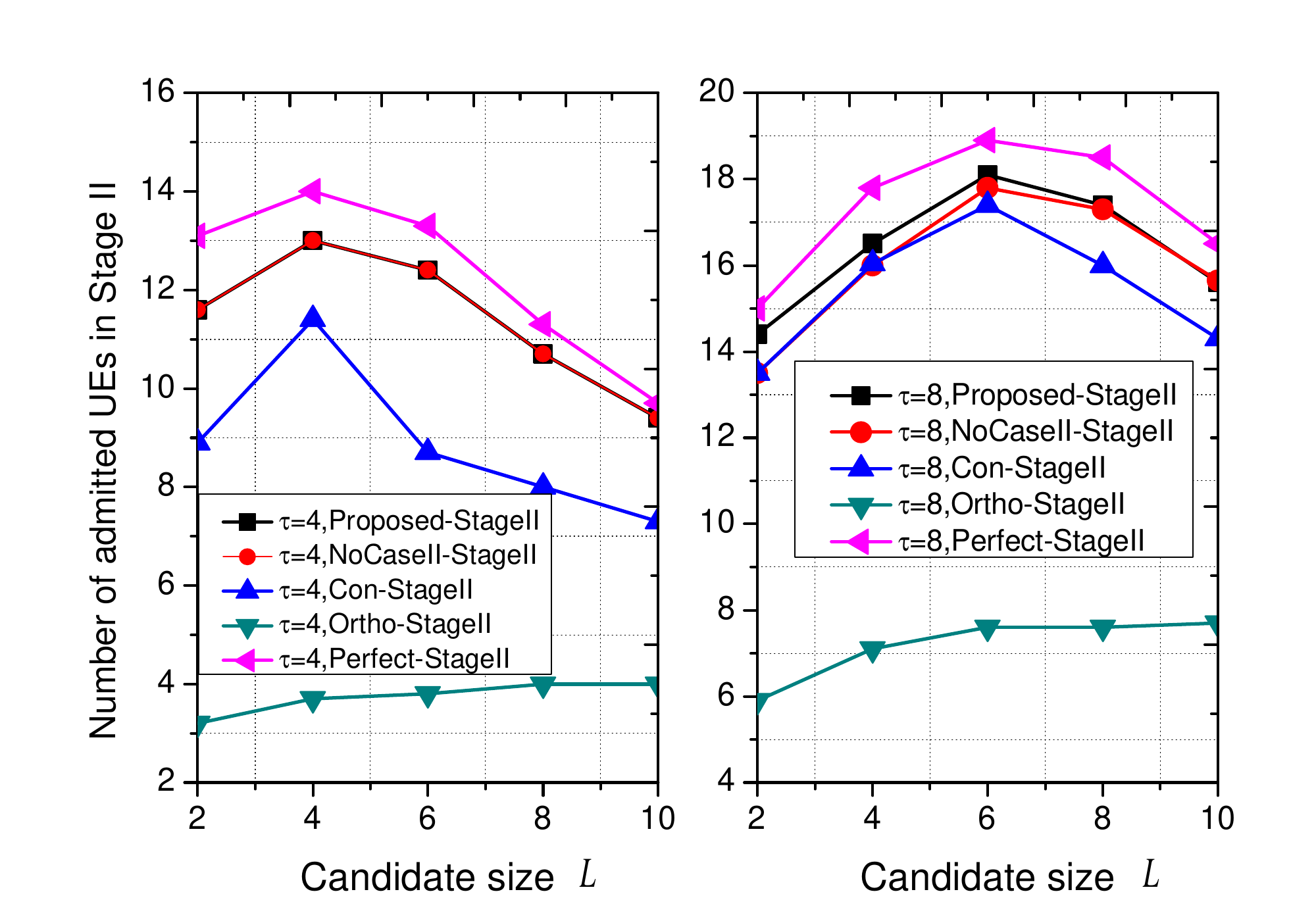}\vspace{-0.4cm}
\caption{Number of admitted UEs in Stage II versus the candidate size $L$.  The left subplot corresponds to the case of $\tau=4$ while the right one is $\tau=8$.}\vspace{-0.8cm}
\label{fig7}\vspace{-0.4cm}
\end{minipage}%
\end{figure}
Figs.~\ref{fig6} and~\ref{fig7} illustrate the number of admitted UEs versus the candidate size $L$ in Stage I and Stage II, respectively.  It is seen from Fig.~\ref{fig6} that the numbers of admitted UEs achieved by all the algorithms in Stage I decrease with the candidate size. The reason is that with the increase of candidate size, more UEs will be connected with each other when constructing the undirected graph. In this case, more UEs will be removed in this stage to satisfy conditions C1 and C2 in Problem ${\cal P}_1$. Fig.~\ref{fig6} also shows that our proposed algorithm achieves superior performance over the ``Con'' algorithm, highlighting the importance of carefully considering pilot interference when removing UEs.

It is interesting to observe from Fig.~\ref{fig7} that the numbers of admitted UEs in Stage II obtained by all the algorithms (except the ``Ortho'' algorithm) initially increase with the candidate size, and then decrease. The reason for the former part is due to the increased spatial degrees of freedom with the increased candidate size. However, when the candidate size continues to increase, many UEs have been prohibited to be admitted due to the pilot allocation in Stage I as seen in Fig.~\ref{fig6}. Hence, in Stage II, even all the selected UEs from Stage I can be admitted, the number is small. This trend is different from most of the existing papers \cite{Ha2016,Lakshmana2016,kim2014,pan2017jsac,pan2017joint}, where the system performance always increases with the candidate size. Hence, the cluster size should be properly optimized and larger cluster size may deteriorate the system performance if channel estimation process is considered. From Fig.~\ref{fig7}, we also observe that our proposed algorithm significantly outperforms the ``Con'' algorithm for both $\tau=4$ and $\tau=8$. On the other hand, when $\tau=4$, both the proposed algorithm and the ``NoCaseII'' algorithm have similar performance; when $\tau=8$, our proposed algorithm performs much better than the ``NoCaseII'' algorithm in the low candidate size regime, i.e., $X<6$, and this performance gap decreases with $X$. The reason is that when the candidate size is small, less number of pilots is required. Hence, it is more likely to fall in Case II in Stage I, which leads to superior performance of the proposed algorithm over the ``NoCaseII'' algorithm due to the additional pilot reallocation step of our proposed algorithm. As expected, the ``Ortho" algorithm has the worst performance since no pilot reuse is allowed.

\subsubsection{Impacts of pilot power}
\begin{figure}
\begin{minipage}[t]{0.475\linewidth}
\centering
\includegraphics[width=3.0in]{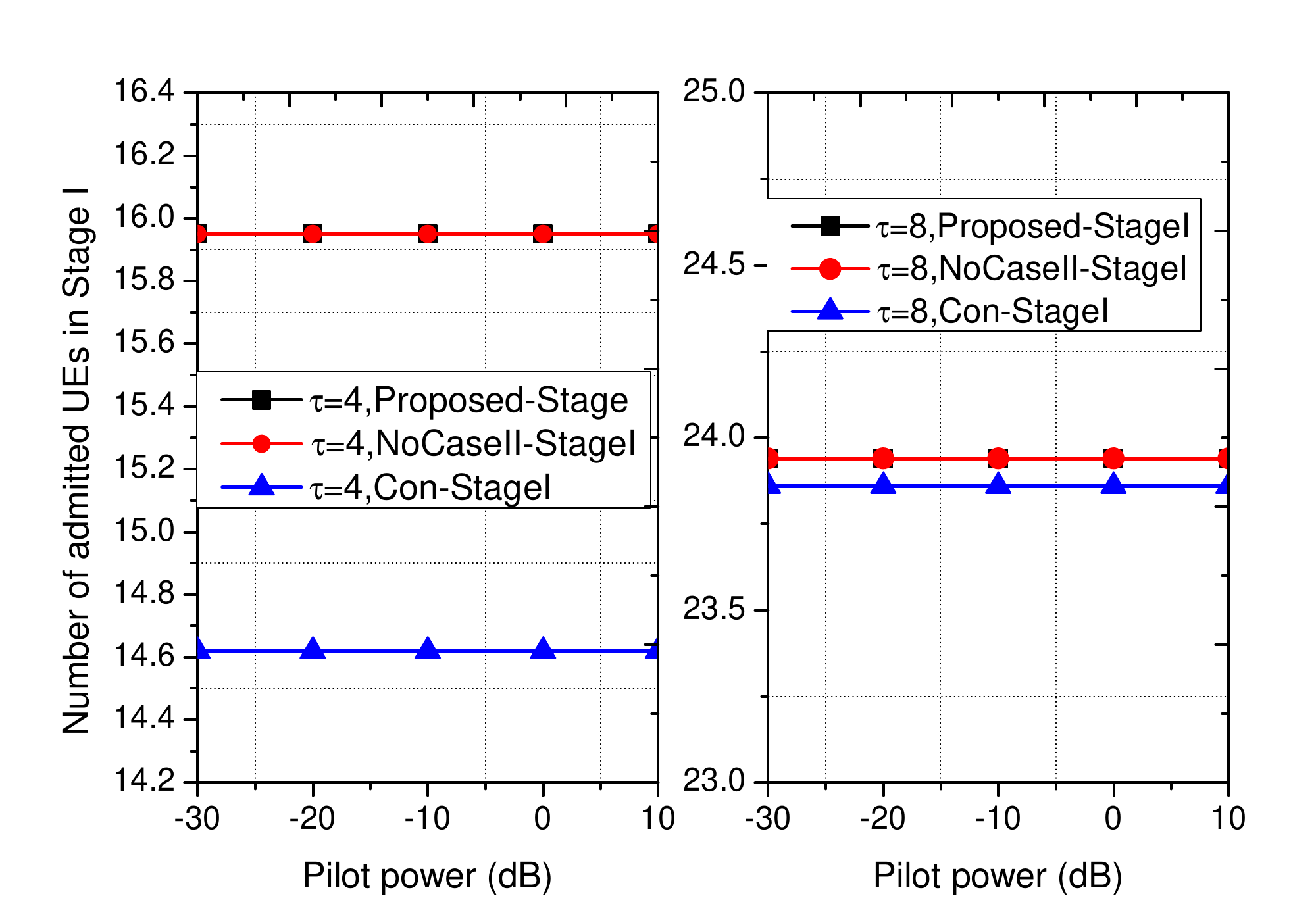}\vspace{-0.4cm}
\caption{Number of admitted UEs in Stage I versus the pilot power. The left subplot corresponds to the case of $\tau=4$ while the right one is $\tau=8$.}\vspace{-0.8cm}
\label{fig8}
\end{minipage}%
\hfill
\begin{minipage}[t]{0.475\linewidth}
\centering
\includegraphics[width=3.0in]{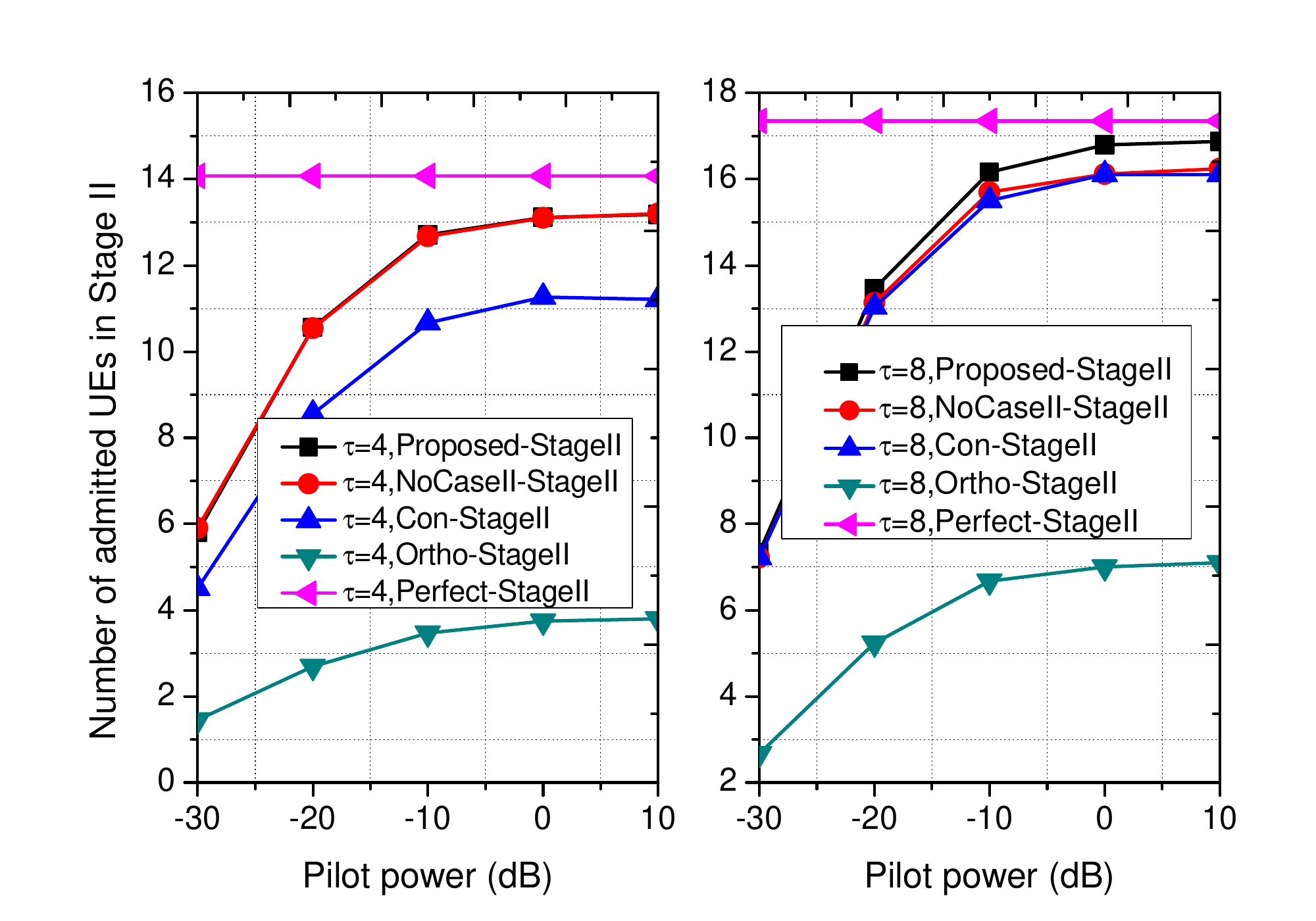}\vspace{-0.4cm}
\caption{Number of admitted UEs in Stage II versus the pilot power.  The left subplot corresponds to the case of  $\tau=4$ while the right one is $\tau=8$.}\vspace{-0.8cm}
\label{fig9}
\end{minipage}%
\end{figure}
Figs.~\ref{fig8} and~\ref{fig9} show the number of admitted UEs versus the pilot power for Stage I and Stage II, respectively. It is again seen from Fig.~\ref{fig8} that our proposed algorithm significantly outperforms the ``Con'' algorithm in Stage I when $\tau=4$, and all algorithms can admit all UEs when $\tau=8$ is Stage I.

Fig.~\ref{fig9} shows that the numbers of admitted UEs obtained by all the algorithms (except the ``Perfect'' algorithm) increase with the pilot power for both cases of $\tau=4$ and $\tau=8$ due to the more accurate CSI. However, there is a fixed gap between the proposed algorithm and the ``Perfect'' algorithm in the high pilot power regime. The reason can be explained as follows. According to the channel estimation error in (\ref{epl}), when the pilot power $p_t$ increases, ${\hat \delta }$ will approach zero, i.e., ${\hat \delta }\rightarrow 0$. Hence, the channel estimation error will become a fixed value, which is only related to the large-scale parameters and the pilot reuse scheme. In this case, the system will enter a pilot interference limited regime. Fig.~\ref{fig9} again demonstrates that the performance superiority of the proposed algorithm over the existing algorithms.

\subsubsection{Impacts of path loss exponent}

\begin{figure}
\begin{minipage}[t]{0.475\linewidth}
\centering
\includegraphics[width=3.0in]{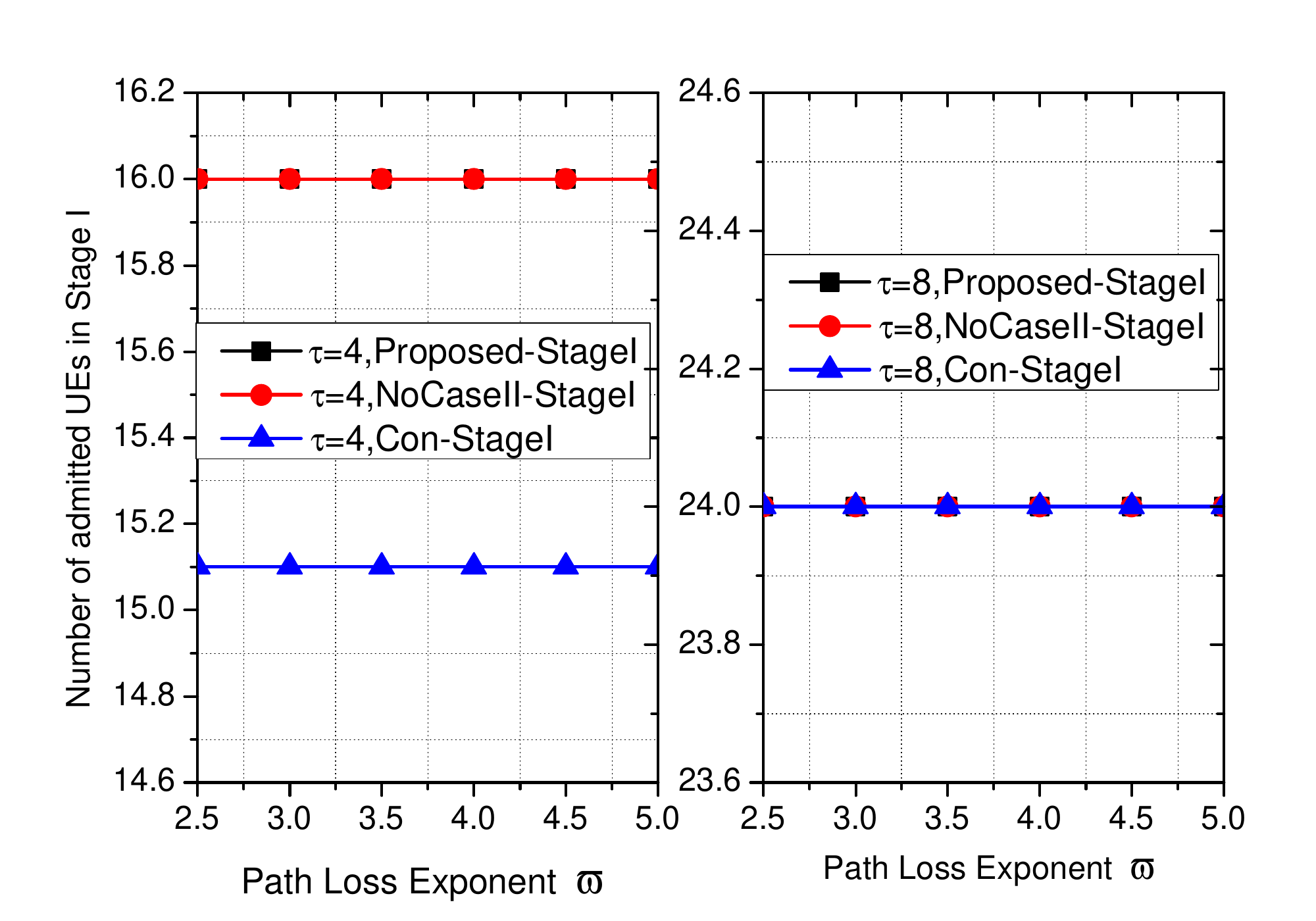}\vspace{-0.4cm}
\caption{Number of admitted UEs in Stage I versus the path loss exponent. The left subplot corresponds to the case of $\tau=4$ while the right one is $\tau=8$.}\vspace{-0.8cm}
\label{figpathstageI}
\end{minipage}%
\hfill
\begin{minipage}[t]{0.475\linewidth}
\centering
\includegraphics[width=3.0in]{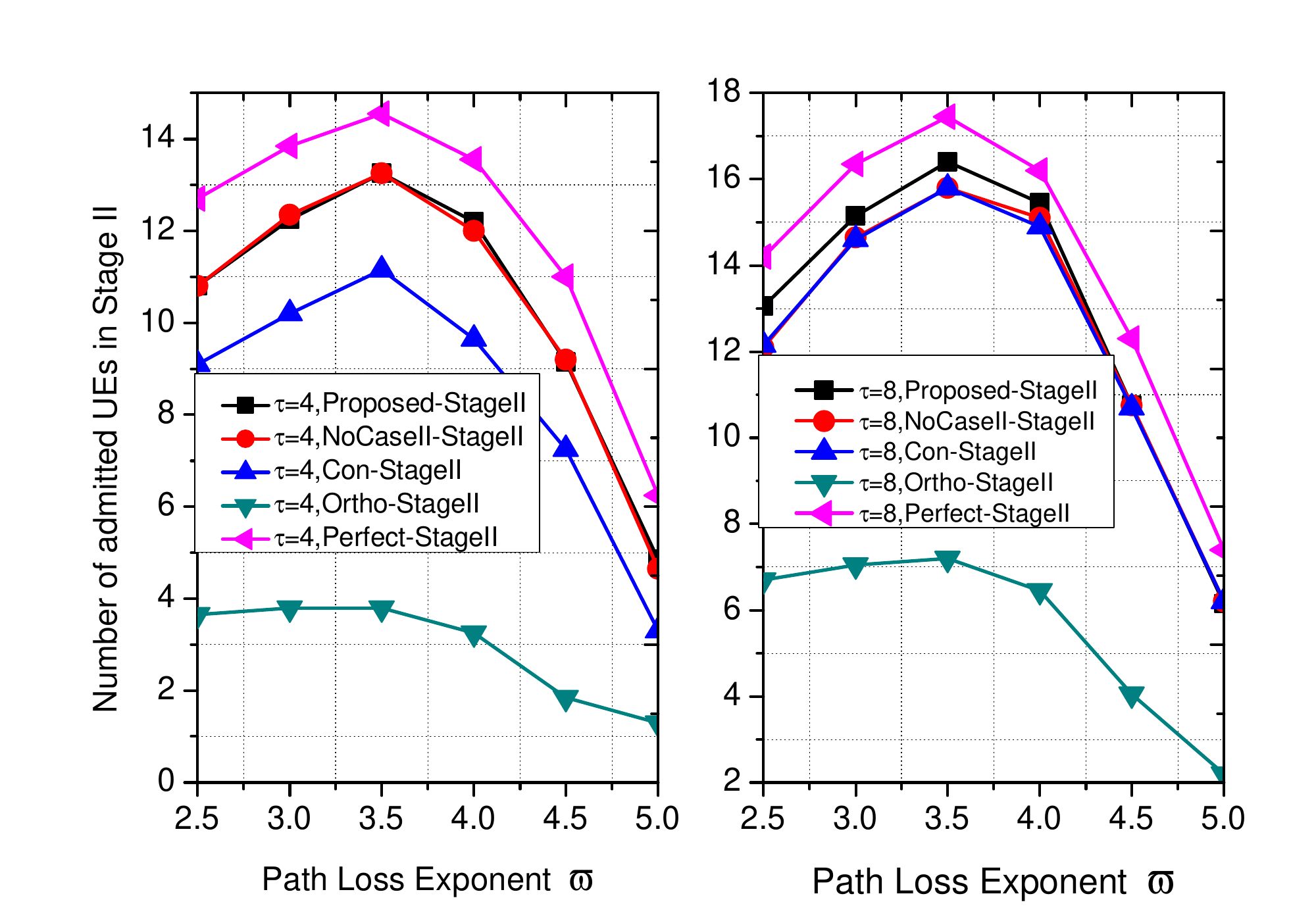}\vspace{-0.4cm}
\caption{Number of admitted UEs in Stage II versus the path loss exponent.  The left subplot corresponds to the case of  $\tau=4$ while the right one is $\tau=8$.}\vspace{-0.8cm}
\label{figpathstageII}
\end{minipage}%
\end{figure}
The impact of path loss exponent is studied here. Specifically, the channel path loss is modeled as $PL = 148.1 + 10 \varpi {\log _{10}}d\ ({\rm{dB}})$, where $\varpi$ denotes the  path loss exponent. Note that the other simulation results in this  section are obtained by setting $\varpi=3.76$. Figs.~\ref{figpathstageI} and~\ref{figpathstageII} show the numbers of admitted UEs versus the path loss exponent $\varpi$ for Stage I and Stage II, respectively. It is observed from Fig.~\ref{figpathstageI} that nearly one more UE can be admitted by the proposed algorithm over the ``Con'' algorithm in Stage I.

It is interesting to find from Fig.~\ref{figpathstageII} that the number of admitted UEs in Stage II for all algorithms (except the ``Ortho'' algorithm) first increases with $\varpi$ and then decreases with it. The reason can be explained as follows. When $\varpi$ is small, the interference power from other RRHs is significant. As a result, the number of UEs that can satisfy their rate targets is lower. With the increase of $\varpi$, the interference power is reduced, and the number of UEs admitted is increasing. However, by additionally increasing $\varpi$, i.e., $\varpi>3.5$, the performance achieved by all algorithms become worse since the  signal power is becoming weaker. Hence, there is an optimal $\varpi$ that all algorithms have the best performance. The performance gain over the existing algorithms is observed in Fig.~\ref{figpathstageII}.

\subsubsection{Impacts of noise power spectral density}
\begin{figure}
\begin{minipage}[t]{0.475\linewidth}
\centering
\includegraphics[width=3.0in]{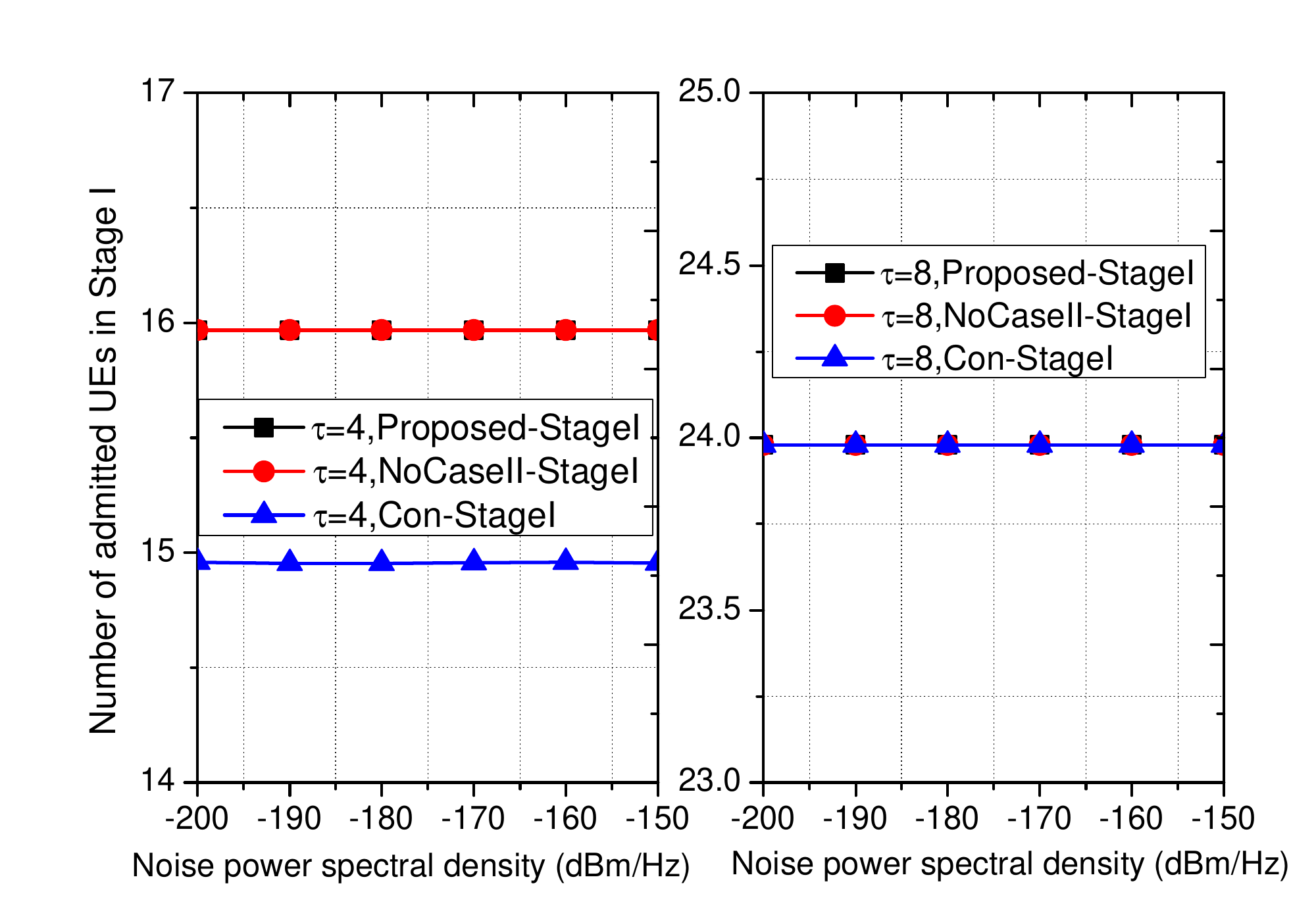}\vspace{-0.4cm}
\caption{Number of admitted UEs in Stage I versus the noise power density. The left subplot corresponds to the case of $\tau=4$ while the right one is $\tau=8$.}\vspace{-0.4cm}
\label{fignoisestageI}
\end{minipage}%
\hfill
\begin{minipage}[t]{0.475\linewidth}
\centering
\includegraphics[width=3.0in]{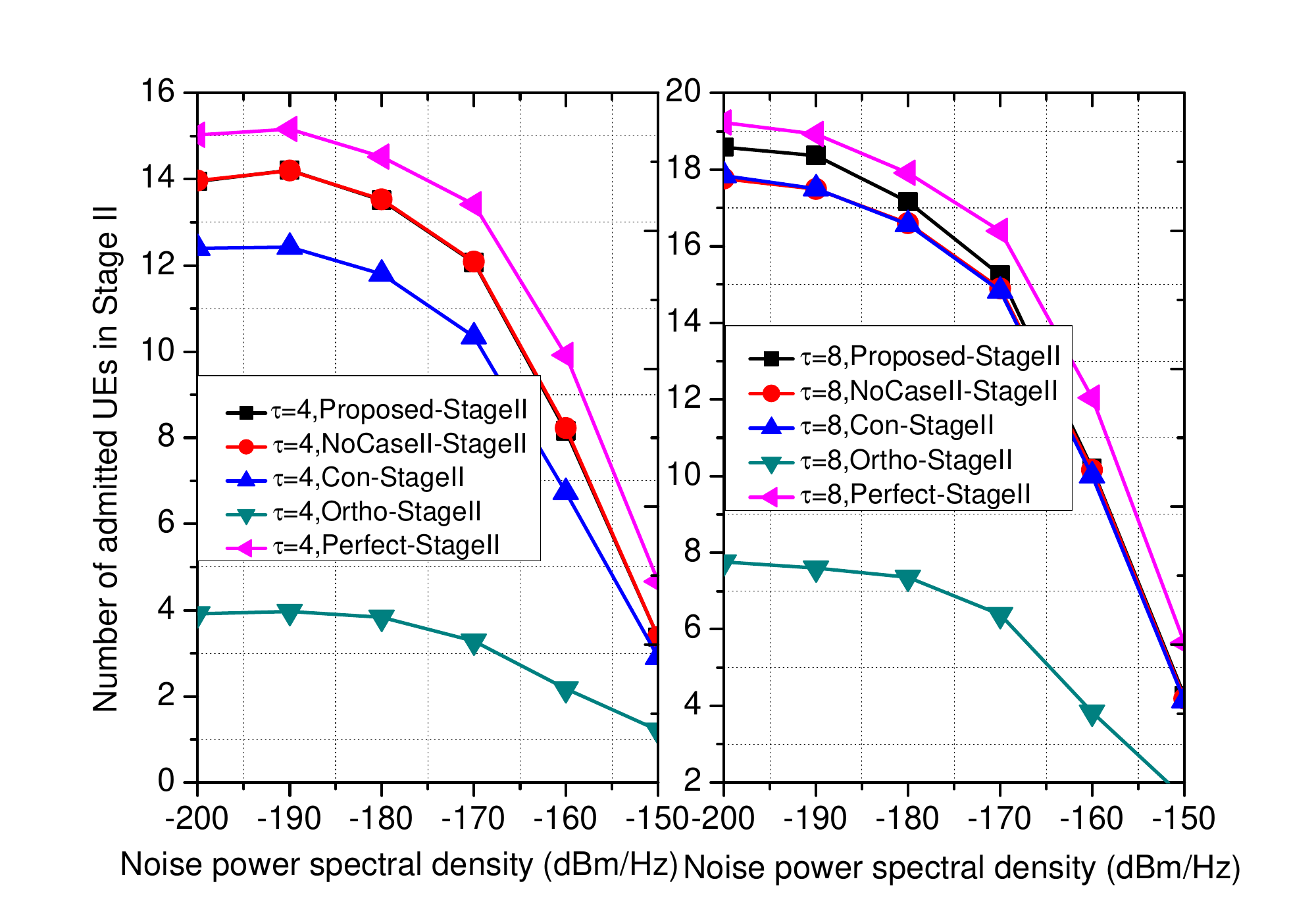}\vspace{-0.4cm}
\caption{Number of admitted UEs in Stage II versus the noise power density.  The left subplot corresponds to the case of  $\tau=4$ while the right one is $\tau=8$.}\vspace{-0.4cm}
\label{fignoisestageII}
\end{minipage}%
\end{figure}
Figs.~\ref{fignoisestageI} and~\ref{fignoisestageII} show the number of admitted UEs versus the noise power density for Stage I and Stage II, respectively.  Fig.~\ref{fignoisestageI} shows the performance superiority of our proposed algorithm over the `Con' algorithm and the number of admitted UEs achieved by all algorithms keep fixed in Stage I.

As expected, from Fig.~\ref{fignoisestageII}, it is observed that the numbers of admitted UEs achieved by all algorithms decrease rapidly with the noise power density in Stage II, and only about five UEs can be admitted by our proposed algorithm when the noise power density is -150 dBm/Hz and $\tau=8$. Fig.~\ref{fignoisestageII} also shows that our proposed algorithm can achieve better performance than the ``Con'' algorithm for both cases of $\tau$, and  than the ``NoCaseII'' algorithm when $\tau=8$.

\subsubsection{Impacts of maximum fronthaul capacity}

\begin{figure}
\begin{minipage}[t]{0.475\linewidth}
\centering
\includegraphics[width=3.0in]{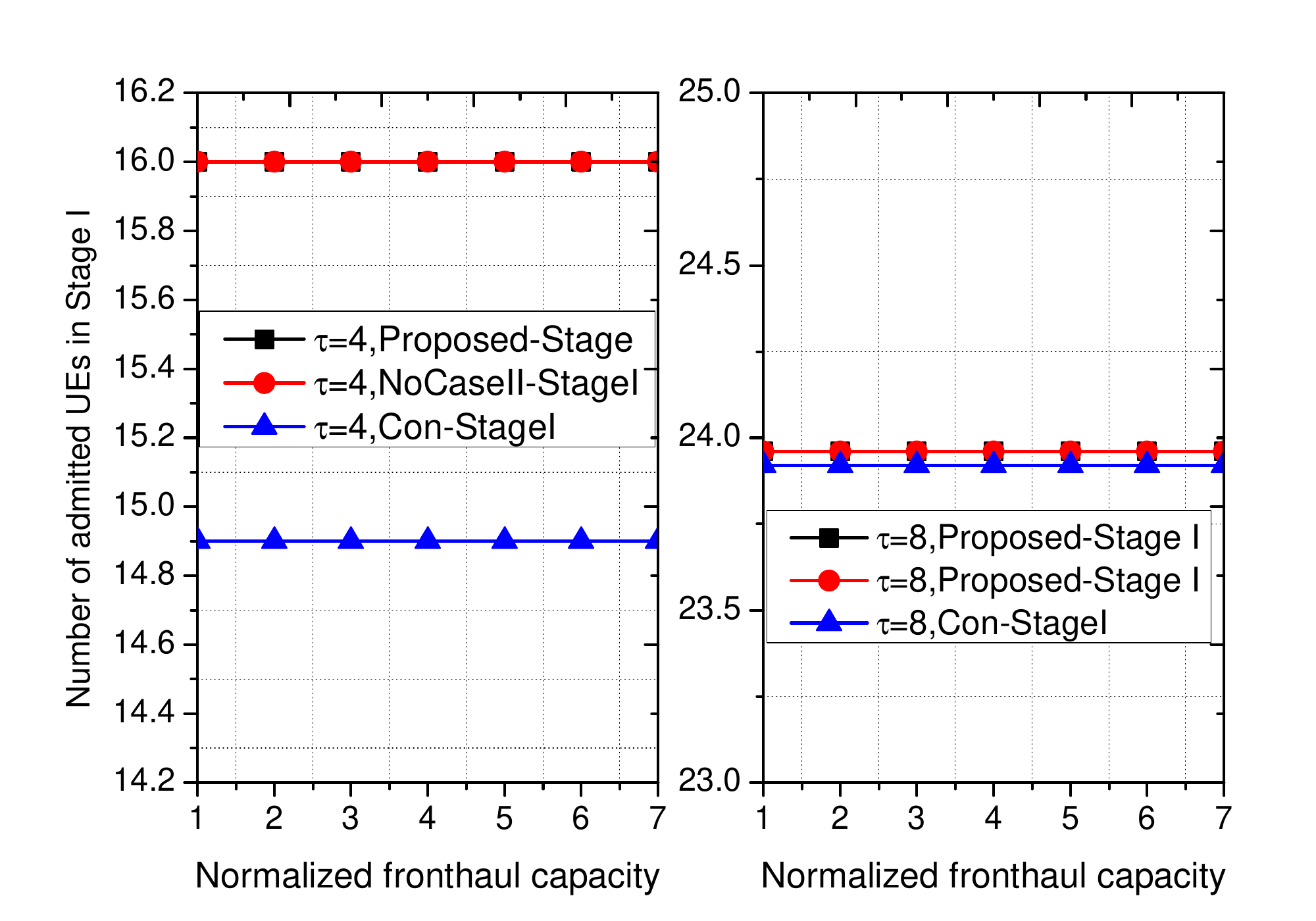}\vspace{-0.4cm}
\caption{Number of admitted UEs in Stage I versus the maximum fronthaul capacity ${\tilde C_{\max }}$. The left subplot corresponds to the case of $\tau=4$ while the right one is $\tau=8$.}
\label{fig12}\vspace{-0.4cm}
\end{minipage}%
\hfill
\begin{minipage}[t]{0.475\linewidth}
\centering
\includegraphics[width=2.85in]{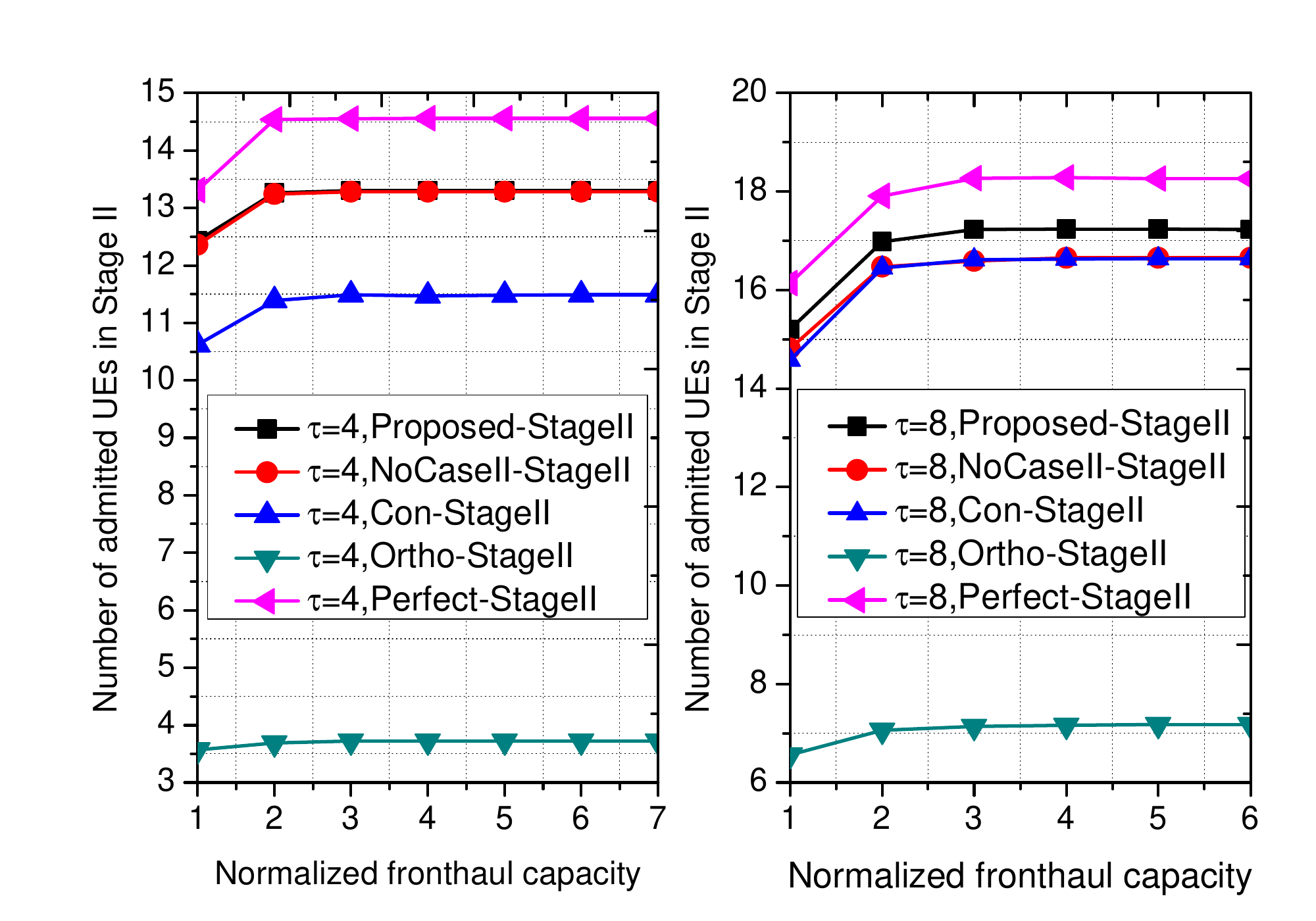}\vspace{-0.4cm}
\caption{Number of admitted UEs in Stage II versus the maximum fronthaul capacity ${\tilde C_{\max }}$.  The left subplot corresponds to the case of $\tau=4$ while the right one is $\tau=8$.}
\label{fig13}\vspace{-0.4cm}
\end{minipage}%
\end{figure}
 Next, we study the impact of fronthaul capacity on the performance of various algorithms. For the sake of illustration, we assume that all UEs have the same rate targets, which are set as ${R_{\min }} = 4{\rm{ bit/s/Hz}}$.  In addition, all fronthaul links are assumed to have the same fronthaul capacity constraints, i.e., ${C_{\max }} = {C_{i,\max }},\forall i$. For ease of exposition, the normalized fronthaul capacity is considered, i.e., ${{\tilde C}_{\max }} = {{{C_{\max }}} \mathord{\left/{\vphantom {{{C_{\max }}} {{R_{\min }}}}} \right. \kern-\nulldelimiterspace} {{R_{\min }}}}$, which denotes the number of UEs that each fronthaul link can support. Figs.~\ref{fig12} and~\ref{fig13} illustrate the number of admitted UEs versus the maximum normalized fronthaul capacity ${\tilde C_{\max }}$ for Stage I and Stage II, respectively. As expected, from Fig.~\ref{fig12}, we can find that all algorithms achieve fixed number of admitted UEs in Stage I, since the pilot allocation process does not depend on ${\tilde C_{\max }}$.

One insightful observation can be found in Fig.~\ref{fig13}: the numbers of admitted UEs by all algorithms initially increase with ${\tilde C_{\max }}$ due to the fact that more UEs can be supported by each fronthaul link. However, when additionally increasing ${\tilde C_{\max }}$, these numbers remain fixed and do not vary much in the large ${\tilde C_{\max }}$ domain for both $\tau=4$ and $\tau=8$. It is shown that ${\tilde C_{\max }}=3$ is sufficient for all algorithms to achieve a large portion of the performance of the case ${\tilde C_{\max }}=\infty$. This observation is instructive for the practical implementation of the dense C-RAN, where the mmWave communication technology with limited bandwidth can be employed as the wireless fronthaul link in dense C-RAN
network.

\subsubsection{Convergence behavior}
\begin{figure}
\centering
\includegraphics[width=3.0in]{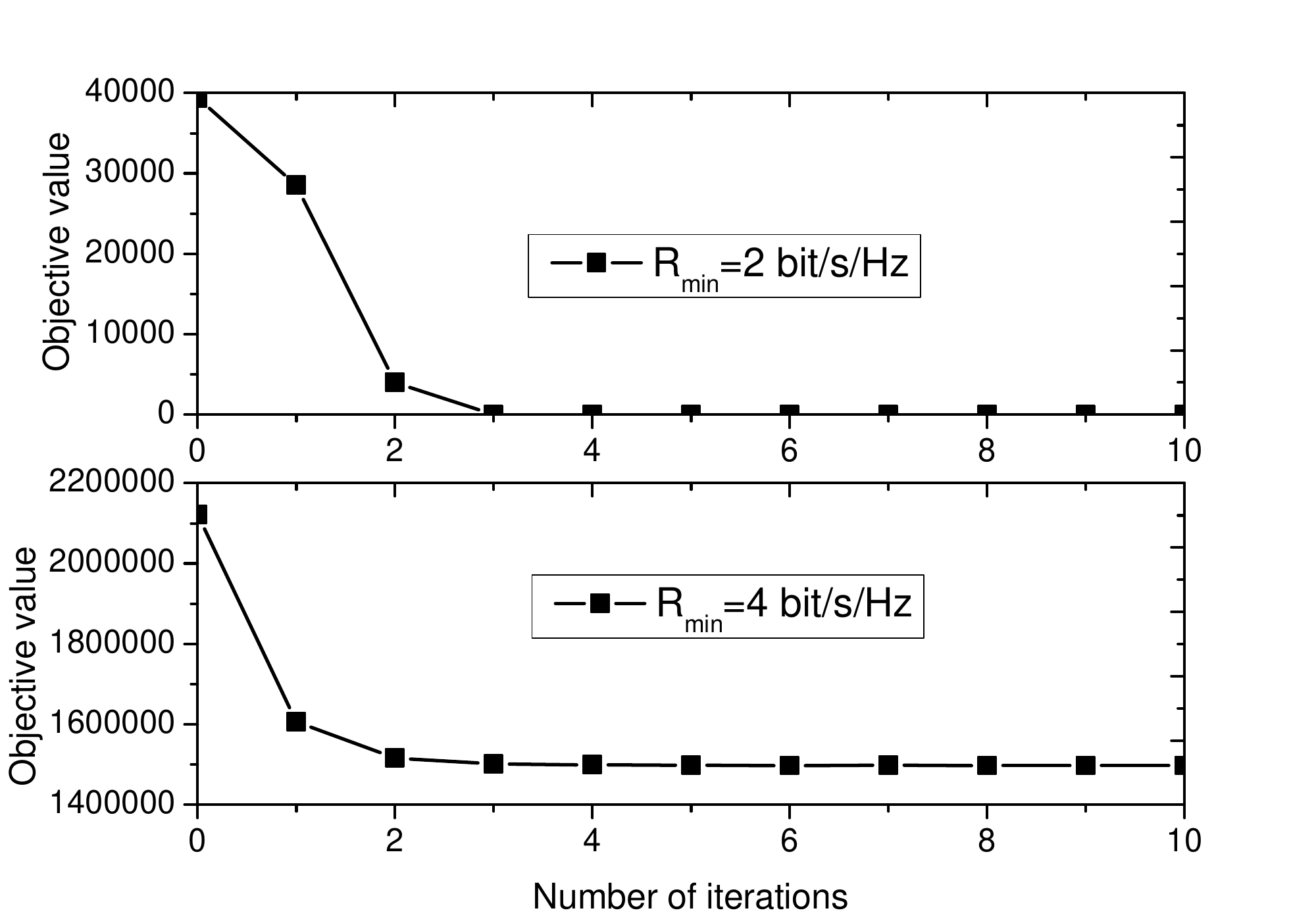}\vspace{-0.4cm}
\caption{Convergence behaviour of the SCA algorithm. The upper subplot shows the feasible case with low rate requirements $R_{\rm{min}}=2\ {\rm{bit/s/Hz}}$, while the lower subplot shows the infeasible case with high rate requirements $R_{\rm{min}}=4\ {\rm{bit/s/Hz}}$.}\vspace{-0.8cm}
\label{fig14}\vspace{-0.4cm}
\end{figure}
Finally, we study the convergence behaviour of the SCA algorithm in Algorithm \ref{algorithmitersca}. Fig.~\ref{fig14} illustrates the objective value of Problem ${\cal P}_6$ versus the number of iterations for two cases of $R_{\rm{min}}=2\ {\rm{bit/s/Hz}}$ and $R_{\rm{min}}=4\ {\rm{bit/s/Hz}}$ for one randomly generated C-RAN network. It is seen from this figure that the objective values for these two cases decrease rapidly and converge within four iterations, which demonstrate the effectiveness and low-complexity associated with our algorithm. When  $R_{\rm{min}}=2\ {\rm{bit/s/Hz}}$, the objective value converges to a very small value, which indicates all UEs can be supported. However, when $R_{\rm{min}}=4\ {\rm{bit/s/Hz}}$, the objective value finally converges to a very large value, which implies the system is infeasible and some UEs should be removed by using our proposed UE selection algorithm in Algorithm \ref{selctalg}.

\vspace{-0.5cm}\section{Conclusions}\label{conclu}
In this paper, we provided a new framework to handle the new challenges of user-centric ultra-dense C-RAN by considering a two-stage problem: the channel estimation for intra-cluster CSI in Stage I and robust transmit beam-vectors design in Stage II. Specifically, in Stage I, we jointly optimized the UE selection and pilot allocation to maximize the number of admitted UEs with limited number of available pilots. One smart UE selection algorithm was proposed when the number of available pilots is not sufficient to support all the UEs. On the other hand, when the opposite case happens, one novel pilot reallocation was introduced to fully use the available pilots. Under the results of Stage I, robust transmit beam-vector was designed to minimize the transmit power while guaranteeing each UE's rate requirement and fronthaul capacity constraints. One novel algorithm was proposed to solve this problem with convergence guarantee. Simulation results verify the effective of the proposed algorithm in terms of the number of admitted UEs compared with the existing naive pilot allocation method. Some interesting observations have been found in the simulations. For example, increasing the cluster/candidate size may not lead to the increased performance when taking the channel estimation into account. Hence, the cluster size should be carefully decided when designing the transmission scheme. In addition, the maximum fronthaul capacity is not necessary to be extremely high and it is shown that ${\tilde C_{\max }}=3$ is sufficient to achieve good performance, which is appealing for the application of mmWave communication link as fronthaul links.

\numberwithin{equation}{section}
\begin{appendices}
\vspace{-0.5cm}\section{Proof of Theorem 1}\label{rateequlity}
We prove this theorem by using the contradiction method. Specifically, suppose one of the following two cases occurs:
\begin{enumerate}
  \item All UEs' data rate lower bounds (LBs) are strictly larger than their rate targets, i.e., $\tilde r_k({\bf{ w}}^\star)>{R_{k,\min }}, \forall k\in \widetilde {\cal U}$;
  \item Part of UEs' data rate LBs are larger than their rate targets (denote this subset of UEs as ${{\widetilde {\cal U}}_1}$), while those of the rest are equal to their targets (denote as this subset as ${{\widetilde {\cal U}}_2}$).
\end{enumerate}

We first prove that the first case cannot happen. To this end, one can construct another feasible solution that yields lower objective value than that of ${\bf{ w}}^\star$. Specifically, we scale each UE's beam-vector by a constant $\sqrt \kappa$ and denote the set of new beam-vectors as ${{\bf{w}}^\# } = \left[ {\sqrt \kappa  {\bf{w}}_k^\star {\rm{,}}k \in \widetilde {\cal U}} \right]$, where $\kappa$ is chosen as
\begin{equation}\label{deaftas}
 \kappa {\rm{ = }}\mathop {\max }\limits_{k \in \widetilde {\cal U}} \frac{{\sigma _k^2}}{{\frac{{{{\left| {{\bf{\hat g}}_{k,k}^{\rm{H}}{\bf{w}}_k^ \star } \right|}^2}}}{{{\eta _{k,\min }}}} - {\bf{w}}_k^{ \star {\rm{H}}}{{\bf{E}}_{k,k}}{\bf{w}}_k^ \star  - \sum\nolimits_{l \ne k,l \in \widetilde {\cal U}} {{\bf{w}}_l^{ \star {\rm{H}}}{{\bf{A}}_{l,k}}{\bf{w}}_l^ \star } }}
\end{equation}
with ${\eta _{k,\min }} = {2^{\frac{{{R_{k,\min }}T}}{{T - \tau }}}} - 1$. It can be readily verified that the constant $\kappa$ is strictly smaller than one, i.e., $\kappa<1$, and all UEs' data rate LBs $\tilde r_k({\bf{ w}}^\#)$ satisfy the following relationships:
\begin{equation}\label{djoieh}
 \tilde r_k({\bf{ w}}^\star)>\tilde r_k({\bf{ w}}^\#)\geq{R_{k,\min }}, \forall k\in \widetilde {\cal U}.
\end{equation}
Since ${\bf{w}}^\star$ is the optimal solution of Problem ${\cal P}_3$, then it must be a feasible solution that satisfies constraints C5, C7 and C8. Hence, the new set of beam-vectors can be verified to be a feasible solution of Problem ${\cal P}_3$, but yields a lower objective value. Hence, this contradicts the assumption that ${\bf{ w}}^\star$ is the optimal solution and  this case cannot happen.

For the second case, we cannot prove it by using the above method directly. The reason is that if $\kappa $ is chosen as in (\ref{deaftas}), its value will be equal to one since some UEs' rate LBs are equal to their targets. To deal with this issue, we divide the group of UEs in $\widetilde {\cal U}$ into two groups, i.e., ${\widetilde {\cal U}_1} = \left\{ {k|{{\tilde r}_k}({{\bf{w}}^ \star }) > {R_{k,\min }},\forall k \in \widetilde {\cal U}} \right\}$ and ${\widetilde {\cal U}_2} = \left\{ {k|{{\tilde r}_k}({{\bf{w}}^ \star }) = {R_{k,\min }},\forall k \in \widetilde {\cal U}} \right\}$. Similar to the first case, we introduce a new constant variable defined as follows:
\begin{equation}\label{deajiru}
 \kappa {\rm{ = }}\mathop {\max }\limits_{k \in \widetilde {\cal U}_1} \frac{{\sigma _k^2}}{{\frac{{{{\left| {{\bf{\hat g}}_{k,k}^{\rm{H}}{\bf{w}}_k^ \star } \right|}^2}}}{{{\eta _{k,\min }}}} - {\bf{w}}_k^{ \star {\rm{H}}}{{\bf{E}}_{k,k}}{\bf{w}}_k^ \star  - \sum\nolimits_{l \ne k,l \in \widetilde {\cal U}} {{\bf{w}}_l^{ \star {\rm{H}}}{{\bf{A}}_{l,k}}{\bf{w}}_l^ \star } }}.
\end{equation}
Then, the constant $\kappa$ is strictly smaller than one. We scale  the beam-vectors of UEs in $\widetilde {\cal U}_1$
 by a constant $\sqrt \kappa$, i.e., ${\bf{w}}_k^\#=\sqrt \kappa {\bf{w}}_k^\star, \forall k \in \widetilde {\cal U}_1$, while keeping the beam-vectors in $\widetilde {\cal U}_2$ unchanged. Denote the set of new beam-vectors as ${{\bf{w}}^\# } = \left[ {\bf{w}}_k^\#,\forall k \in \widetilde {\cal U}_1, {  {\bf{w}}_k^\star {\rm{,}}k \in \widetilde {\cal U}_2} \right]$. It can be verified the following relations hold:
 \begin{equation}\label{djoreawa}
 \tilde r_k({\bf{ w}}^\star)>\tilde r_k({\bf{ w}}^\#)\ge{R_{k,\min }}, \forall k\in \widetilde {\cal U}_1,
\end{equation}
and
\begin{equation}\label{djoreawa}
\tilde r_k({\bf{ w}}^\#)> \tilde r_k({\bf{ w}}^\star)={R_{k,\min }}, \forall k\in \widetilde {\cal U}_2.
\end{equation}
Since the data rate LBs in $\widetilde{\cal U}_2$ increase, the fronthaul capacity constraints in C8 may not hold. To resolve this problem, we update the new UE sets $\widetilde {\cal U}_1$ and $\widetilde{\cal U}_2$, and repeat the above procedure. Since the new constructed beam-vectors decrease the total power consumption and the objective value is lower bounded. Hence, this procedure will converge to a fixed value. Note that if there exists some UEs whose rate LBs are strictly larger than their targets, we can  always construct new beam-vectors with reduced total power. Hence, when the procedure converges, all UEs data rates are equal to their targets, and the final beam-vectors satisfy all the constraints C5, C7 and C8. However, the objective value with the final beam-vectors have a lower value than that with ${\bf{ w}}^\star$, which contradicts the assumption that ${\bf{ w}}^\star$ is the optimal solution. Hence, this case cannot happen either, and the proof is complete.

\vspace{-0.5cm}\section{Proof of Theorem 2}\label{prooftheorem1}
Denote the collection of $\{{\bf{W}}_k,\forall k\}$ as ${\bf{W}}$. Then, the Lagrangian function of Problem ${\cal P}_9$ is
\begin{eqnarray}
 &&{\cal L}\left( {{\bf{W}},{\bf{x}},\bm \lambda,\bm \mu,\bm \nu, \bm \omega } \right) = \sum\limits_{k \in {\cal U}} {{\rm{tr}}\left( {{{\bf{W}}_k}} \right)}  + \sum\limits_{k \in  {\cal U}} {x_k}  + \sum\limits_{i \in {\cal I}} {{\lambda _i}}{\left( {\sum\limits_{k \in {{\cal U}_i}} {{\rm{tr}}\left( {{{\bf{W}}_k}{{\bf{B}}_{i,k}}} \right)}  - {P_{i,\max }}} \right)}+\nonumber\\
 &&\quad\sum\limits_{k \in  {\cal U}} {{\mu _k}\left( {{\eta _{k,\min }}\left( {{\rm{tr}}\left( {{{\bf{W}}_k}{{\bf{E}}_{k,k}}} \right) + \sum\limits_{l \ne k,l \in {\cal U}} {{\rm{tr}}\left( {{{\bf{W}}_l}{{\bf{A}}_{l,k}}} \right)}  + \sigma _k^2} \right) - {\rm{tr}}\left( {{{\bf{W}}_k}{{{\bf{\hat g}}}_{k,k}}{\bf{\hat g}}_{k,k}^{\rm{H}}} \right) - {x_k}} \right)},\nonumber \\
&&\quad +\sum\limits_{i \in {\cal I}} {{\nu _i}\left( {\sum\limits_{k \in {{\cal U}_i}} {{\tau _{i,k}}{\rm{tr}}\left( {{{\bf{W}}_k}{{\bf{B}}_{i,k}}} \right)}  - {{\tilde C}_i}} \right)} - \sum\limits_{k \in  {\cal U}} {{\rm{tr}}\left( {{{\bf{W}}_k}{{\bf{Z}}_k}} \right)}-\sum\limits_{k \in {\cal U}} {{\omega _k}{x_k}} ,\nonumber
 \end{eqnarray}
where $\bm \lambda  = \left[ {{\lambda _i},\forall i \in {\cal I}} \right]$, $\bm \mu  = \left[ {{\mu _k},\forall k \in  {\cal U}} \right]$, and $\bm \nu  = \left[ {{\nu _i},\forall i \in {\cal I}} \right]$ are  the non-negative Lagrangian multipliers associated with constraints C13, C14 and C15, respectively. $\bm \omega=\left[ {{\omega _k},\forall k \in {\cal U}} \right] $ are the dual variables associated with the nonnegative constraints of ${\bf{x}}=\{x_k, \forall k \in {\cal U}\}$, and $\{{{{\bf{Z}}_k}}\succeq {\bf{0}},\forall k \in {\cal U}\}$ denote the dual variable matrices associated with the semi-definite constraints on $\{{\bf{W}}_k,\forall k\in {\cal U}\}$.

As Problem ${\cal P}_9$ is a convex optimization problem, its globally optimal solution must satisfy its first-order optimality condition as:
\begin{eqnarray}
&&\frac{{\partial {\cal L}\left( {{\bf{W}},\lambda ,\mu ,\nu } \right)}}{{\partial {{\bf{W}}_k}}} = {{\bf{I}}_{M\left| {{{\cal I}_k}} \right| \times M\left| {{{\cal I}_k}} \right|}} + \sum\limits_{i \in {\cal I}} {\left( {{\lambda _i} + {\nu _i}{\tau _{i,k}}} \right){{\bf{B}}_{i,k}}}  - {\mu _k}{{{\bf{\hat g}}}_{k,k}}{\bf{\hat g}}_{k,k}^{\rm{H}} + \nonumber\\
&&\qquad\qquad\quad \sum\limits_{k \in  {\cal U}} {{\mu _k}{\eta _{k,\min }}{{\bf{E}}_{k,k}}}  + \sum\limits_{l \ne k,l \in  {\cal U}} {{\mu _l}{\eta _{l,\min }}{{\bf{A}}_{k,l}}}  - {{\bf{Z}}_k}  = {\bf{0}}, \forall k\in {\cal U}.
 \end{eqnarray}
Then, ${{\bf{Z}}_k}$ can be represented as ${{\bf{Z}}_k} = {{\bf{D}}_k} - {\mu _k}{{{\bf{\hat g}}}_{k,k}}{\bf{\hat g}}_{k,k}^{\rm{H}}$, where ${{\bf{D}}_k}$ is given by
\begin{equation}\label{defineD}
  {{\bf{D}}_k} = {{\bf{I}}_{M\left| {{{\cal I}_k}} \right| \times M\left| {{{\cal I}_k}} \right|}} + \sum\limits_{i \in {\cal I}} {\left( {{\lambda _i} + {\nu _i}{\tau _{i,k}}} \right){{\bf{B}}_{i,k}}}  + \sum\limits_{k \in {\cal U}} {{\mu _k}{\eta _{k,\min }}{{\bf{E}}_{k,k}}}  + \sum\limits_{l \ne k,l \in {\cal U}} {{\mu _l}{\eta _{l,\min }}{{\bf{A}}_{k,l}}}.
\end{equation}
According to the Karush-Kuhn-Tucker (KKT) condition, we have
\begin{eqnarray}
&&{{\mu _k}\!\left(\! {{\eta _{k,\min }}\left(\! {{\rm{tr}}\left( {{{\bf{W}}_k}{{\bf{E}}_{k,k}}} \right)\! +\!\! \sum\limits_{l \ne k,l \in {\cal U}} {{\rm{tr}}\left( \!{{{\bf{W}}_l}{{\bf{A}}_{l,k}}} \!\right)} \! +\! \sigma _k^2} \right) \!\!-\!\! {\rm{tr}}\left(\! {{{\bf{W}}_k}{{{\bf{\hat g}}}_{k,k}}{\bf{\hat g}}_{k,k}^{\rm{H}}}\! \right) - {x _k}}\! \right)}\!=\!0,\forall k,\label{KKTONE}\\
&&{{{\bf{W}}_k}{{\bf{Z}}_k}{ = {\bf{0}}}},\forall k. \label{KKTTWO}
\end{eqnarray}

Before proving the theorem, we first give the following lemma.

\itshape \textbf{Lemma 1:}  \upshape  The optimal Lagrangian dual multipliers $\{{\mu _k},\forall k\}$ are positive, i.e., ${\mu _k}>0, \forall k$.

\itshape \textbf{Proof:}  \upshape This can be proved by using contradiction. Denote the optimal solution of Problem ${\cal P}_9$ as $\left\{ {{{\bf{W}}_k^\star},{x_k^\star},\forall k} \right\}$ and the corresponding Lagrangian multipliers as $\left[ {{\lambda _i^\star},{\nu _i^\star}, {\mu _k^\star}, \forall i \in {\cal I}, \forall k \in {\cal U} } \right]$.  Assume there exists one ${\mu _l^\star}$ that is zero. Then, according to KKT conditions in (\ref{KKTONE}), we have
${\rm{tr}}\left( {{{\bf{W}}_l^\star}{{{\bf{\hat g}}}_{l,l}}{\bf{\hat g}}_{l,l}^{\rm{H}}} \right) + {x _l^\star}> {\eta _{l,\min }}\left( {{\rm{tr}}\left( {{{\bf{W}}_l^\star}{{\bf{E}}_{l,l}}} \right) + \sum\nolimits_{m\ne l,m \in {\cal U}} {{\rm{tr}}\left( {{{\bf{W}}_m^\star}{{\bf{A}}_{m,l}}} \right) + \sigma _l^2} } \right)$. We consider the following two cases:

1) If ${x _l^\star}>0$, we can find a new ${{\bar x }_l} = x _l^ \star  - \Delta $, where $\Delta>0$ is a small enough positive value such that ${\rm{tr}}\left( {{{\bf{W}}_l^\star}{{{\bf{\hat g}}}_{l,l}}{\bf{\hat g}}_{l,l}^{\rm{H}}} \right) + {\bar x _l} \geq{\eta _{l,\min }}\left( {{\rm{tr}}\left( {{{\bf{W}}_l^\star}{{\bf{E}}_{l,l}}} \right) + \sum\nolimits_{m \ne l,m \in {\cal U}} {{\rm{tr}}\left( {{{\bf{W}}_m^\star}{{\bf{A}}_{m,l}}} \right) + \sigma _l^2} } \right)$ holds. Then, we find another feasible solution $\left\{ {{\bf{W}}_k^ \star ,\forall k,x_k^ \star ,\forall k \ne l,{{\bar x }_l}} \right\}$ that yields a lower objective value than that of $\left\{ {{{\bf{W}}_k^\star},{x _k^\star},\forall k} \right\}$, which contradicts that $\left\{ {{{\bf{W}}_k^\star},{x _k^\star},\forall k} \right\}$ is the optimal solution.

2) If ${x _l^\star}=0$, we can find a new precoding matrix ${{{\bf{\bar W}}}_l} = \rho {{\bf{W}}_l^\star}$ with $0<\rho<1$ such that ${\rm{tr}}\left( {{{\bf{\bar W}}_l}{{{\bf{\hat g}}}_{l,l}}{\bf{\hat g}}_{l,l}^{\rm{H}}} \right)\geq {\eta _{l,\min }}\left( {{\rm{tr}}\left( {{{\bf{\bar W}}_l}{{\bf{E}}_{l,l}}} \right) + \sum\nolimits_{m\ne l,m \in {\cal U}} {{\rm{tr}}\left( {{{\bf{W}}_m^\star}{{\bf{A}}_{m,l}}} \right) + \sigma _l^2} } \right)$ holds since ${{\bf{A}}_{m,l}},\forall m \ne l$ are positive definite matrices. Then, we find a new feasible solution $\left\{ {{\bf{W}}_k^ \star ,\forall k \ne l,{{{\bf{\bar W}}}_l},x _k^ \star ,\forall k} \right\}$ that yields a lower objective value than that of $\left\{ {{{\bf{W}}_k^\star},{x _k^\star},\forall k} \right\}$, which contradicts that $\left\{ {{{\bf{W}}_k^\star},{x _k^\star},\forall k} \right\}$ is the optimal solution.
 \hfill $\Box$

The first term of ${{\bf{D}}_k}$ in (\ref{defineD}) is an identity matrix. In addition, the Lagrange multipliers $\{ {{\lambda _i},{\nu _i}, {\mu _k}, \forall i,k } \}$ and $\{{\tau _{i,k}},\forall i,k\}$ are nonnegative values, and $\{{\bf{B}}_{i,k},{{\bf{E}}_{k,k}},{{\bf{A}}_{k,l}}\}$ are positive definite matrices. As a result, ${{\bf{D}}_k}$ is a positive definite matrix with ${\rm{rank}}({{\bf{D}}_k})=M\left| {{{\cal I}_k}} \right|$. Since ${\rm{rank(}}{{\bf{Z}}_k}{\rm{)}} \ge {\rm{rank}}({{\bf{D}}_k}) - {\rm{rank}}({\mu _k}{{{\bf{\hat g}}}_{k,k}}{\bf{\hat g}}_{k,k}^{\rm{H}})$, we obtain ${\rm{rank(}}{{\bf{Z}}_k}{\rm{)}}\ge M\left| {{{\cal I}_k}} \right|-1$, where we use the fact that ${\rm{rank}}({\mu _k}{{{\bf{\hat g}}}_{k,k}}{\bf{\hat g}}_{k,k}^{\rm{H}})=1$ since ${\mu _k}$ is positive according to Lemma 1. Furthermore, according to (\ref{KKTTWO}), we have ${\rm{rank}}({{\bf{W}}_k}) \le M\left| {{{\cal I}_k}} \right| - {\rm{rank(}}{{\bf{Z}}_k}{\rm{)}}$. Then, we have ${\rm{rank}}({{\bf{W}}_k}) \le 1$. Obviously, the optimal precoder ${\bf{W}}_k$ is not a zero matrix with ${\rm{rank}}({{\bf{W}}_k})=0 $. Hence, the optimal precoder  ${\bf{W}}_k$ has rank one, i.e., ${\rm{rank}}({{\bf{W}}_k})=1 $, which completes the proof.

\end{appendices}

\
\



%\bibliographystyle{IEEEtran}
%% argument is your BibTeX string definitions and bibliography database(s)
%\bibliography{myre}



\vspace{-1.2cm}
\bibliographystyle{IEEEtran}
\bibliography{myref}


\end{document}